\documentclass[onecolumn]{aa}
\usepackage{txfonts}
\usepackage{geometry}                
\usepackage{rotating}
\usepackage{natbib}
\usepackage{graphicx}
\usepackage{amssymb}
\usepackage{color}
\usepackage{epstopdf}
\bibpunct{(}{)}{;}{a}{}{,}

\graphicspath{{figures/}}

\begin{document}
\title{Stability of quasi-Keplerian shear flow in a laboratory experiment}
\author{Ethan Schartman\inst{1}\thanks{Current address: Nova Photonics, Inc. Princeton, NJ, USA}
\and Hantao Ji\inst{1}\thanks{All correspondence should be addressed to hji@pppl.gov}
\and Michael J. Burin\inst{2}\thanks{Current address: Department of Physics, California State University San Marcos, San Marcos, CA , USA}
\and Jeremy Goodman\inst{2}
}
\institute{Princeton Plasma Physics Laboratory, Princeton University, Princeton, NJ, USA 
\and Department of Astrophysical Sciences, Princeton University, Princeton, NJ, USA
}
\date{\today}                                           
\abstract{Subcritical transition to turbulence has been proposed as a source of turbulent viscosity required for the associated angular momentum transport for fast accretion in Keplerian disks. Previously cited laboratory experiments in supporting this hypothesis were performed either in a different type of flow than Keplerian or without quantitative measurements of angular momentum transport and mean flow profile, and all of them appear to suffer from Ekman effects, secondary flows induced by nonoptimal axial boundary conditions. Such Ekman effects are expected to be absent from astronomical disks, which probably have stress-free vertical boundaries unless strongly magnetized.}
{To quantify angular momentum transport due to subcritical hydrodynamic turbulence, if exists, in a quasi-Keplerian flow with minimized Ekman effects. }
{We perform a local measurement of the azimuthal--radial component of the Reynolds stress tensor in a novel laboratory apparatus where Ekman effects are minimized by flexible control of axial boundary conditions.} 
{We find significant Ekman effects on angular momentum transport due to nonoptimal axial boundary conditions in quasi-Keplerian flows. With the optimal control of Ekman effects, no statistically meaningful angular momentum transport is detected in such flows at Reynolds number up to two millions.} 
{Either a subcritical transition does not occur, or, if a subcritical transition does occur, the associated radial transport of angular momentum in optimized quasi-Keplerian laboratory flows is too small to directly support the hypothesis that subcritical hydrodynamic turbulence is responsible for accretion in astrophysical disks. Possible limitations in applying laboratory results to astrophysical disks due to experimental geometry are discussed. }
\keywords{accretion, accretion disks -- hydrodynamics -- instabilities -- turbulence}
\maketitle

\section{Introduction}

The everyday experience of velocity shear in a fluid flow is that if the shear is large enough a laminar flow will become turbulent. In non-rotating flow such as Poiseiulle flow or plane-Couette flow the laminar state is linearly stable at all Reynolds numbers. Nevertheless, in the presence of  disturbances of finite amplitude, a transition to turbulence can occur~\citep{Darbyshire:1994}. 
In the case of rotating shear flow, such as the Taylor-Couette flow ~\citep{Taylor:1923,Couette:1888} established between concentric rotating cylinders, a transition to turbulence was observed~\citep{Wendt:1933,Taylor:1936} for the case in which the inner cylinder was at rest with the outer one spinning.  This configuration is linearly stable to infinitesimal perturbations at all Reynolds numbers. Nevertheless, instability was observed at $Re\sim 10^3-10^5$. The transition has been assumed to be subcritical due to similarities with pipe flow transition, though    \citet{Schultzgrunow:1959} stated that the transition appeared to be similar to a Kelvin-Helmholtz instability. No first-principles theory exists for subcritical transition in a rotating shear flow, though recent efforts to identify such a transition mechanism can be found in \citet{Chagelishvili:2003, Tevzadze:2003, Yecko:2004, Umurhan:2004, Mukhopadhyay:2005, Afshordi:2005, Lithwick07, Rincon:2007, Lithwick09, Mukhopadhyay:2011a, Mukhopadhyay:2011b}. Phenomenological models have been developed based on the assumption that the observed transition is subcritical in nature~\citep{Zeldovich:1981,Richard:1999,Longaretti:2002,Dubrulle:2005}. 

This observation has motivated hydrodynamic theories of accretion mechanisms since the work of \cite{Shakura:1973}. The linear MagnetoRotational Instability (MRI)~\citep{Balbus:1991} has since emerged as the mechanism most likely to provide the required levels of angular momentum transport. However, the electrical conductivity of protostellar accretion disks is at best marginally adequate for MRI-driven accretion \citep{Bai:2011,Oishi:2011} and references therein. Therefore, purely hydrodynamic models based on subcritical hydrodynamic instability (SHI) still remain of interest for these cool disks.

A review of transport mechanisms in accretion disks can be found in \cite{Balbus:1998}.
The matter in thin accretion disks follow Keplerian angular velocity profiles, $\Omega(r)\propto r^{-3/2}$, the angular momentum increases with radius, $l(r) \propto r^2\Omega \propto r^{1/2}$ and the disk is stable to linear centrifugal instability~\citep{Rayleigh:1916}.
In SHI models of disk accretion, shear in the angular velocity is the source of free energy to drive enhanced transport. A dimensionless coefficient $\beta$ relates the shear to a turbulent viscosity, $\nu_T$.
Modeling by \citet{Hueso:2005} of two protoplanetary systems require that 
$\beta$ lies in the range $10^{-6}-10^{-4}$, which appears to be compatible with 
the deduced $\beta$ by \citet{Richard:1999} based on the Taylor-Couette flow experiments by \citet{Wendt:1933} and \citet{Taylor:1936}.

The velocity profile developed  in a Taylor-Couette experiment may share with the Keplerian flow the opposing gradients of angular momentum  and angular velocity. We refer to these flows as \emph{quasi-Keplerian}. The quasi-Keplerian profile is a subset of \emph{anti-cyclonic} profiles in which  $\partial \Omega/\partial r  < 0$ and  may be either centrifugally unstable, $\partial l/\partial r<0$,  or  stable, $\partial l/\partial r>0$. Flows in which $\partial \Omega/\partial r  > 0$ are \emph{cyclonic} and are linearly stable at all Reynolds numbers. This is the regime in which Wendt and Taylor made their observations. It is worth noting that the $\beta$ prescription by \citet{Richard:1999}, even if valid for the cyclonic evidence on which they based it, may not be valid for anti-cyclonic flow, in particular Keplerian flow.

\begin{figure}[htbp] 
   \centering
   \includegraphics[width=3in]{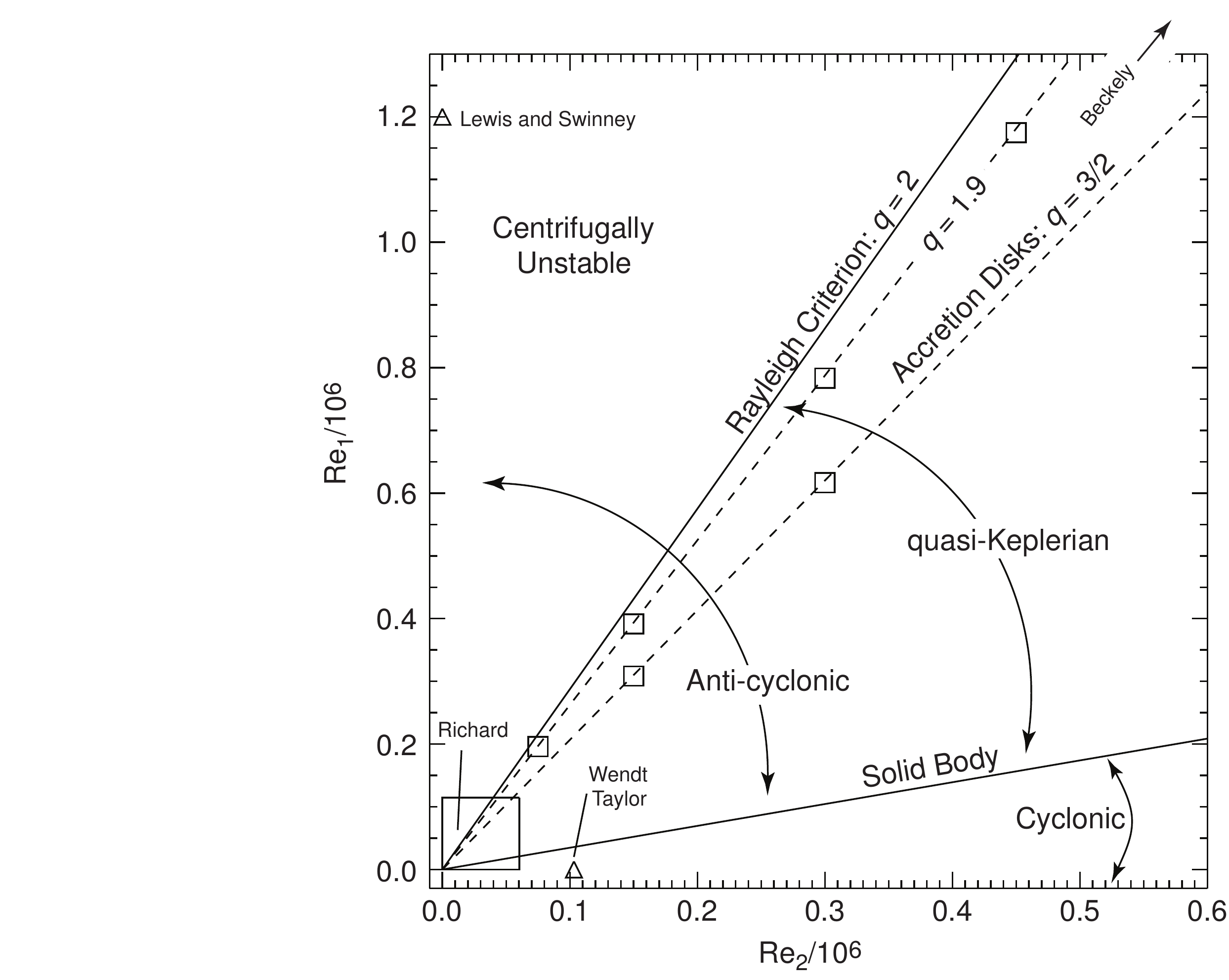} 
   \caption{Rotation profiles developed between concentric, co-rotating cylinders. The axes are the rotation speeds of the cylinders defined by a Reynolds number $Re_{1,2} = \Omega_{1,2}r_{1,2}(r_2-r_1)/\nu$, where 1 (2) refers to the inner (outer) cylinder. The cylinder radius is $r$ and $\Omega$ is the angular velocity. The kinematic viscosity is $\nu$. The flow profiles are classified by a local exponent of angular velocity $q\equiv-\partial\ln \Omega/\partial\ln r$. In cyclonic flow $q<0$. The anti-cyclonic regime is further divided in to  quasi-Keplerian and centrifugally unstable flows, the boundary between the two is the flat angular momentum profile, $q=2$.  }
   \label{fig:re_params}
\end{figure}

The Princeton MRI experiment is a Taylor-Couette device developed~\citep{Schartman:2009} to produce quasi-Keplerian flows in both water and a liquid gallium alloy to look for evidence of both SHI using water and the MRI using alloy~\citep{Ji:2001, Goodman:2002}. 
 The flows of interest are diagrammed in Figure~\ref{fig:re_params}. We use two dimensionless numbers to characterize our flow profiles: the Reynolds number, $Re$, and the local exponent of angular velocity, $q$. The concentric  cylinders have radii $r_1, r_2~(r_1<r_2)$ and angular velocities $\Omega_1,\Omega_2$. We define 
  $Re \equiv \left(\Omega_1 -\Omega_2 \right)\left(r_2-r_1\right)\bar{r}/\nu$, where $\bar{r} = (r_2+r_1)/2$ and $\nu$ is the kinematic viscosity. We follow the definition used in \citet{Balbus:1998} for $q\equiv-\partial\ln\Omega/\partial\ln r$. For a Keplerian disk $q=3/2$, and the limit of marginal stability by the Rayleigh criterion occurs for $q=2$. 

Based on observations of SHI in non-rotating flow three characteristics are always present. 1) The laminar state before transition is linearly stable to infinitesimal perturbations. 2) A hysteresis is observed in the transition: the change from the laminar to the turbulent state occurs at a greater Reynolds number than the transition from turbulent to laminar flow. 3) The transition to the turbulent state is triggered by the presence of ambient fluctuations in the laminar flow. For circular Poiseuille flow no transition occurs below  $Re\approx 2\times 10^3$, even for perturbation amplitude equal  to the mean flow velocity~\citep{Darbyshire:1994}. At higher Reynolds number, the  fluctuation level required to trigger transition decreases. In carefully controlled experiments the transition has been delayed  to $Re\sim10^5$~\citep{Pfenniger:1961}.  

Local shearing box simulations have been performed to investigate the hydrodynamic stability of quasi-Keplerian flows. \citet{Hawley:1999} and \citet{Lesur:2005} studied the decay of applied turbulence near the boundary of marginal centrifugal instability~\citep{Rayleigh:1916}.  
Both investigations excite transiently  the largest Fourier modes of the simulation domain and observe the subsequent evolution of the flow. In both simulations  the turbulence decays rapidly for all quasi-Keplerian values of $q$, except those close to centrifugal instability.  \citet{Hawley:1999} report that for $q/2 < 99.5\%$ and perturbation amplitudes of ${v'/ v}=1~\mathrm{to}~100\%$  the flow relaminarizes within one to five orbital periods. 

In the anti-cyclonic cases investigated by \citet{Lesur:2005} they begin the simulation on the marginally-centrifugally stable boundary, $q=2$, with a perturbation amplitude of 100\%. While holding $Re$ constant, they gradually step in to the centrifugally stable regime until the flow becomes laminar. They observe that the turbulence persists into the stable regime, but that the critical Reynolds number is an extremely steep function of the distance to the linear stability boundary. In going from $q/2\approx99\%$ to $q/2\approx97\%$, the transition Reynolds number increases from $<1\times10^4$ to $\approx8\times10^4$.

The forcing amplitude imposed on a fully developed flow used by \citet{Lesur:2005} cannot be produced in the Princeton MRI Experiment. We are restricted to forcing the flow with ambient fluctuation levels generated by the boundaries, or by accelerating the boundaries to produce a new flow profile. In the latter case, a statistically steady mean velocity is not reached before the fluctuation levels fall below 5\%. In the most unstable flows which we can produce via instability, the maximum measured fluctuation amplitude is of order 10\%. No apparent hystereses associated with a transition were observed with these attempts.

Prior to our work in this regime, \citet{Richard:2001} and \citet{Beckley:2002} were the only experiments to study the stability of quasi-Keplerian flows. Richard's experiments were performed at $Re\sim10^4$, and Beckley's were at $Re\sim 10^6$. The relation of these Reynolds numbers to our experiments are shown in Figure~\ref{fig:re_params}. Beckley diagnosed his flow by measuring the torque required to maintain a steady state flow. He found that the Reynolds number scaling of this torque was inconsistent with laminar Taylor-Couette flow. Beckley's apparatus was constructed with cylinder heights comparable to the gap between them, $h = 2(r_2-r_1)$, and the end caps fixed to the outer cylinder. Because of the relative proximity of the end caps and their speed, Beckley concluded that the excess torque was due to an Ekman circulation (see Section~\ref{sec:boundary_flow}). 

The Taylor-Couette device employed by \citet{Richard:2001} used taller cylinders, $h\approx 25(r_2-r_1)$, with less influence on the bulk flow from the end caps than Beckley's apparatus (by using a \lq\lq split" configuration; see Section 3 below). Using a visualization technique, Richard observed that his flow underwent a transition to wavy states~\citep{Longaretti:2008} at a Reynolds number of approximately $10^4$, but without quantitative measurements of the angular momentum transport. The measured profiles of the mean azimuthal velocity, however, still deviated significantly from the ideal Couette solution (see Section 2 below), presumably due to Ekman circulation driven by the endcaps.

In this paper we present the results of our investigation of SHI in quasi-Keplerian flows in the Princeton MRI experiment. We show significant influences by end caps on the measured angular momentum transport. By using two independently rotatable end caps at each end, the Ekman effects are minimized, resulting in much lower transport levels than proposed by \citet{Richard:1999}. The implications to angular momentum transport in astrophysical disks are discussed.

\section{Measurement of angular momentum transport }


We briefly review the hydrodynamics of the ideal Couette flow to make explicit the terms which are accessible experimentally to us. We work in cylindrical  coordinates  $(r,\theta, z)$. 
The incompressible  Navier-Stokes equations with stress tensor $\bar{\bf \sigma}$ are:
\begin{eqnarray}
\frac{\partial {\bf v}}{\partial t} + \left( {\bf v}\cdot\nabla\right ) {\bf v} =& \nabla\cdot{\bar{\bf \sigma}}\\
\nabla \cdot {\bf v} =& 0
\end{eqnarray}
Writing out the full equation  for the azimuthal velocity, $v_\theta$:
\begin{equation}
\label{eqn:NS_theta}
\frac{\partial v_\theta}{\partial t} + {v_r\over r}\frac{\partial (r v_\theta)}{\partial r} + \frac{v_\theta}{r}\frac{\partial v_\theta}{\partial \theta} + v_z\frac{\partial v_\theta}{\partial z} =
-\frac{1}{\rho r}\frac{\partial p}{\partial \theta} + \nu\left( {1\over r^2}{\partial \over \partial r}\left[ r^3 {\partial \over \partial r} \left( v_\theta \over r \right) \right]+\frac{1}{r^2}\frac{\partial^2 v_\theta}{\partial \theta^2} + \frac{\partial^2 v_\theta}{\partial z^2}  + \frac{2}{r^2}\frac{\partial v_r}{\partial \theta} \right)
\end{equation}

For steady-state, laminar flow where the transport is only radial 
 (\emph{i.e.} infinitely tall cylinders), $\partial/\partial\theta = \partial/\partial z = 0$, the incompressibility condition gives $(1/r) \partial( rv_r)/\partial r = 0$ and therefore $v_r =0$ from the impermeable radial boundary conditions. Equation~\ref{eqn:NS_theta} becomes:
\begin{equation}
0 = {1\over r^2}{\partial \over \partial r}\left[ r^3 {\partial \over \partial r} \left( v_\theta \over r \right) \right]
\end{equation}
leading to the ideal Couette solution for the angular velocity:
\begin{equation}\label{eqn:ideal_couette}
\Omega(r) = a + \frac{b}{r^2},  
\quad \mathrm{with~constants:}
\qquad a = \frac{\Omega_2r^2_2 - \Omega_1r_1^2}{r_2^2 - r_1^2}, \qquad b=\frac{\left(\Omega_1-\Omega_2\right)r_2^2r_1^2}{r_2^2 - r_1^2},
\end{equation}
where $\Omega_1~(\Omega_2)$ is the angular velocity of the inner (outer) cylinder, and $r_1~(r_2)$ is the radius of the inner (outer) cylinder, $r_1 < r_2$.
The constants $a, b$ are determined from  non-slip boundary conditions at the cylinder walls. 

We split all quantities into values time-averaged over a period greater than $ \tau_E$ (denoted by $\langle\rangle$) and fluctuations around them (denoted by primes), as in $v_i=\langle v_i \rangle + v_i^\prime$. By the virtue of a rotating flow, time averaging is equivalent to azimuthal averaging so that
$\partial \langle v_i \rangle /\partial t  = \partial \langle v_i \rangle /\partial \theta  = 0$.
The averaged Equation~\ref{eqn:NS_theta} then becomes
\begin{equation}
\label{eqn:NS_theta_avg}
\frac{\langle v_r \rangle}{r} \frac{\partial (r \langle v_\theta\rangle)}{\partial r}  +  \langle v_z\rangle \frac{\partial \langle v_\theta\rangle}{\partial z}+ \frac{1}{r^2}\frac{\partial (r^2 \langle v_r^\prime v_\theta^\prime \rangle)}{\partial r} + \frac{\partial \langle{v_z^\prime  v_\theta^\prime }\rangle}{\partial z} =  \nu\left( {1\over r^2}{\partial \over \partial r}\left[ r^3 {\partial \over \partial r} \left( \langle v_\theta \rangle \over r \right) \right] + \frac{\partial^2 \langle v_\theta \rangle}{\partial z^2} \right).
\end{equation}
The second term on the left hand side here (also on the right hand side) can be dropped because of Taylor-Proudman theorem~\citep{Greenspan:1968}
which states that under steady-state, inviscid conditions where $\partial (r^2\Omega(r))/\partial r>0$ the flow profile becomes independent of height.
This has been confirmed in our experiments (see Fig.\ref{fig:vert_scan} later and also \citet{Ji:2006}), if sufficiently far from the boundaries.
Therefore, deviation from the ideal Couette solution, Equation~\ref{eqn:ideal_couette}, arises from only three sources: advection of momentum by a mean radial velocity $\langle v_r\rangle$, the radial-azimuthal component of the Reynolds stress, $\Sigma_{r\theta} = \langle v_r^\prime v_\theta^\prime \rangle$, and the axial-azimuthal component of Reynolds stress, $\Sigma_{\theta z} =  \langle{v_z^\prime v_\theta^\prime}\rangle$. Using 2-component Laser Doppler Velocimetry (LDV) oriented axially we can directly measure $\Sigma_{r\theta}$ and $\langle v_r\rangle$. We do not  measure $\Sigma_{\theta z}$ because the measurement volumes for $v_\theta$ and $v_z$ do not overlap when the diagnostic is oriented radially~\citep{Schartman:2009}. This is not of any consequence for this study, however, since in accretion disks $\Sigma_{r\theta}$ is the turbulent stress responsible for the enhanced transport and closure models developed for disks focus on it exclusively.




\cite{Zeldovich:1981}  first  applied Taylor's torque measurements of cyclonic Couette flow to develop a closure model for  $\Sigma_{r\theta}$. 
\citet{Richard:1999} extended Zeldovich's analysis using the combined results of Wendt and Taylor to develop their closure model. We will focus here on the methodology of Richard and Zahn, but note that
\citet{Longaretti:2002} arrived at a similar result using an eddy viscosity argument applied to the cyclonic data.

The closure model of Richard and Zahn proceeds from the scaling of the transition Reynolds number in the cyclonic data of   Taylor and  Wendt.  The critical Reynolds number, $Re_c$, is a function of the gap between the cylinders $\Delta r = r_2-r_1$.
When the ratio of gap width to the average radius $\bar{r}= (r_1+r_2)/2$,   $\Delta r/\bar{r} > 1/20$, $Re_c$ increases like $(\Delta r/\bar{r})^2$. In this regime Richard and Zahn rewrite the critical Reynolds number in terms of the angular velocity gradient and arrive at an instability condition:
\begin{equation}\label{eqn:Re_c}
Re_c = \frac{\bar{r}^3}{\nu}\frac{\Delta\Omega}{\Delta r} \left(\frac{\Delta r}{\bar{r}}\right)^2 \ge 6\times 10^5\left(\frac{\Delta r}{\bar{r}}\right)^2.
\end{equation}
(We note that this prescription implies that the relevant length scale is $\bar r$ rather than other length scales, $\Delta r$ or $h$, an assumption which may not be justified. See Section 5 for more discussion.)

By this prescription, \citet{Richard:2001} should expect a turbulent transition at $Re_c\approx7\times10^4$ for the radius ratio $\Delta r/\bar{r} = 0.35$. In the cyclonic experiments, a transition was observed at $Re=3\times10^4$. However, the estimate for $Re_c$ does not carry over to the anti-cyclonic experiment  where a transition was observed at $Re\approx3\times10^3$.
Also absent from the discussion leading  to this equation is an estimate of the perturbation amplitudes which triggered the transitions in the experiments of Wendt and Taylor. 

Torque measurements in the turbulent regime suggest to Richard and Zahn that the turbulent viscosity, $\nu_t$, is a diffusive process and choose for it the form $\nu_t = \alpha \bar{r}\Delta \Omega\Delta r$, $\alpha$ is a constant. 
Finally, using the observed approximate scaling of $\alpha$ with gap width, Richard and Zahn conclude that the  local value for $\nu_t$ becomes independent of the gap width (for large enough gaps) and is  determined only by the local shear:
\begin{equation}\label{eqn:nu_turb}
\nu_t = \beta \left | r^3 \frac{\partial\Omega}{\partial r} \right |.
\end{equation}
The flux of angular momentum is then given by
\begin{equation}
\rho r^2 \langle v_r^\prime v_{\theta}^\prime \rangle = - \rho\nu_{t}r^3\frac{\partial\Omega}{\partial r},
\end{equation}
which can be rewritten in terms of $q$:
\begin{equation}\label{eqn:beta_measure}
\beta = -\langle v_r^\prime v_{\theta}^\prime \rangle / q^2v_{\theta}^2.
\end{equation}
Thus,  $\beta$ can be directly determined through measurements of the Reynolds stress.
Finally, we comment on a particular case where $\langle v_\theta\rangle$ satisfies the ideal Couette solution (Equation \ref{eqn:ideal_couette}) with negligible $\langle v_r\rangle$ and axial dependences. In such a case, Equation \ref{eqn:NS_theta_avg} reduces to
\begin{equation}
\frac{1}{r^2}\frac{\partial (r^2 \langle v_r^\prime v_\theta^\prime \rangle)}{\partial r}=0
\end{equation}
where the specific angular momentum flux, $r^2 \langle v_r^\prime v_{\theta}^\prime \rangle$ [$=\nu_t r^3q\Omega = \beta (r^3q\Omega)^2$], is a spatial constant. 
This is especially convenient when diagnostic access for Reynolds stress measurements are limited to certain locations. The Reynolds stress at other locations can be inferred.


\section{Experiment\label{sec:experiment}}

\begin{figure}[htbp] 
   \centering
   \includegraphics[width=\textwidth]{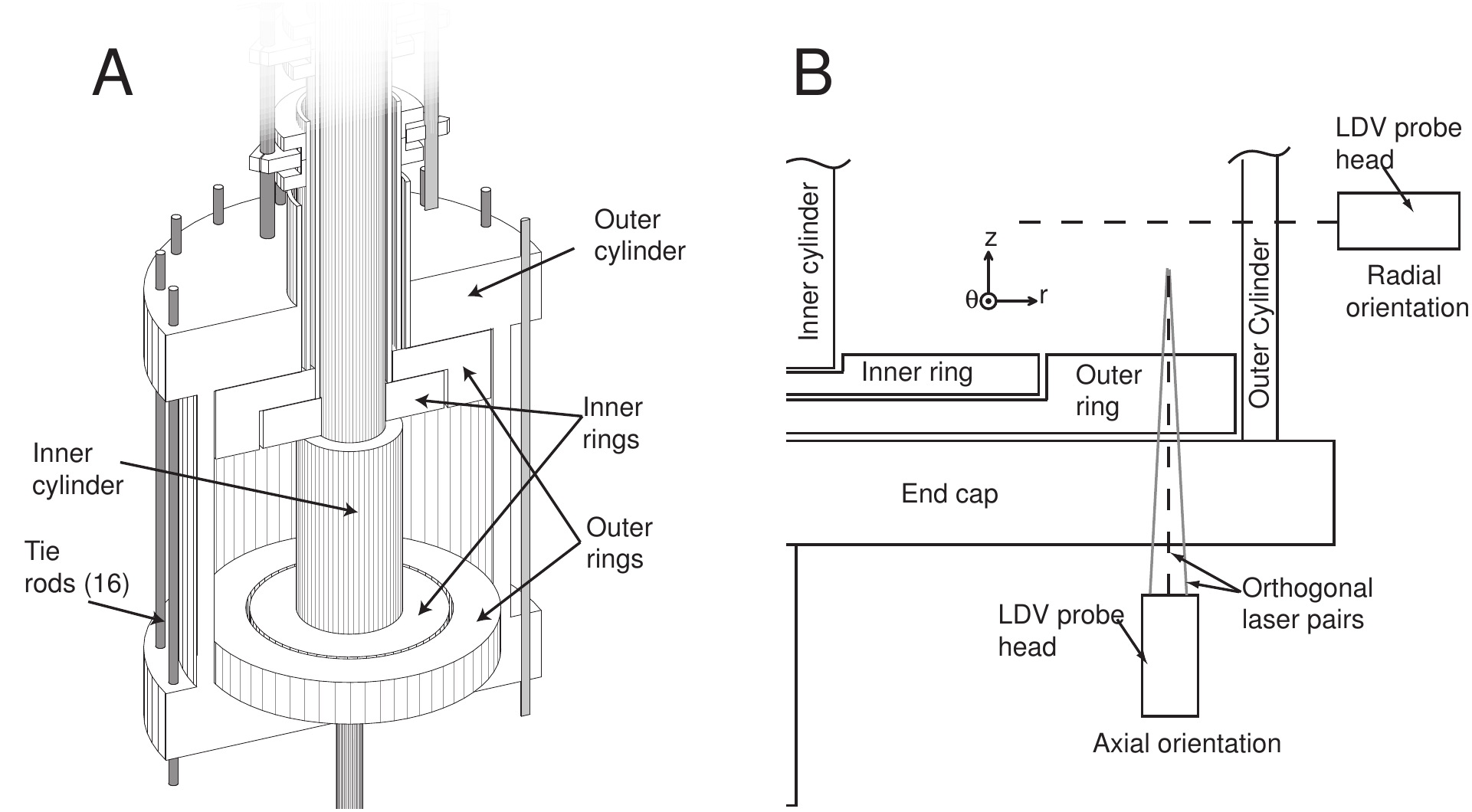} 
   \caption{A. The  Princeton MRI Experiment is a Taylor-Couette apparatus in which the end caps of the vessel have been divided in to two pairs of nested, independently rotatable end rings. B. The primary diagnostic of the fluid velocity is Laser Doppler Velocimetry (LDV). Radial profiles of  $v_\theta$ are acquired with the diagnostic viewing the fluid from the side. In the axial orientation two components of velocity $v_\theta$ and $v_r$ are measured. Simultaneous measurement of the two velocities gives a direct measure of the $r-\theta$ component of the Reynolds stress. Due to very low data rates, the Reynolds stress is measured primarily at one point.}
   \label{fig:apparatus}
\end{figure}

The Princeton MagnetoRotational Instability Experiment is a novel Taylor-Couette apparatus~\citep{Tagg:1994}. The working fluid is confined between concentric, corotating cylinders which are bounded vertically by two pairs of nested, differentially rotating end rings~\citep{Burin:2006,Schartman:2009}, Figure~\ref{fig:apparatus}.
The experiment was designed to produce quasi-Keplerian flows of a liquid gallium alloy, GaInSn~\citep{Morley:2008} which would become unstable to the MagnetoRotational Instability in the presence of an applied solenoidal magnetic field \citep{Ji:2001}.   To minimize the volume of GaInSn required for the MRI studies, the height of the cylinders is only twice the gap width between them, $h/\Delta r= h/\left(r_2 - r_1\right) \approx 2$. 
This aspect ratio  is small in comparison with other Taylor-Couette experiments which aim  to minimize the influence of the end caps by separating  them as much as possible.  For example, \cite{Taylor:1936} used  $h/\Delta r>100$. 
Further details of the design and implementation of the apparatus and diagnostics can be found in \cite{Schartman:2009}.

\subsection{Apparatus}

The experiment  outer cylinder is a pressure vessel  into which the inner cylinder and end rings are submerged. The outer cylinder is a 25.4~mm thick annulus of cast acrylic  capped by two 101.6~mm thick acrylic disks. 
The inner cylinder and  end rings are mounted to nested stainless steel axles which pass through the  top cap of the outer cylinder. The rings are acrylic. The lower rings and outer cylinder cap were polished to allow optical diagnostic access to the fluid.  The inner cylinder is stainless steel and was painted black to reduce reflections which would interfere with the velocity measurement. A lip seal is mounted at the top end of each axle to seal against its inner neighbor. The submerged, lower, end of each component is fixed radially by a plain bearing. 

The hydrodynamic  experiments reported here use water or a water-glycerol mix as the working fluid. The kinematic viscosity of the water-glycerol mix is $\nu_{mix} = (15\pm1) \nu_{water}$, with a density of 1.2~$\mathrm{gm/cm^3}$.

The Princeton MRI Experiment also differs from previous Taylor-Couette experiments in the low tolerance of several components. The submerged plain bearings have radial "play" ranging up to 0.1~ mm. Because those bearings are nested, they contribute to a total run-out of the inner cylinder of up to 0.8~mm, see \cite{Schartman:2009}.

\begin{table}[htbp]
   \centering
   \begin{tabular}{ |c|c c| } 
\hline
      Component &  radius & $\Omega_{max}/2\pi$ \\
 		 & (mm)& ($\mathrm{s}^{-1}$)\\
\hline
	Inner cylinder	& 70.6	& 66.7  \\
	Inner ring 	& 101.5 &  24.3 \\
	Outer ring  	& 167.8 &  6.7\\
	Outer cylinder	& 203.0 & 9.0 \\	
\hline
	Cylinder height &\multicolumn{2}{c|}{280.0~mm}\\
\hline
   \end{tabular}
   \caption{Radii and maximum angular velocities of experiment components. For the inner and outer rings the given radius is the mid-radius of the ring. }
   \label{tab:mech_table}
\end{table}

\subsection{Experimental flow profiles }

The component speeds used to develop the azimuthal velocity profiles are enumerated in Table~\ref{tab:speed_ratios}. The four component speeds are listed as ratios to the outer cylinder rotation speed. Our flows of primary interest are the two quasi-Keplerian profiles with nominal values based on the cylinder speeds  of $q=1.5$ and 1.9. The flow shear of the former is most analogous to accretion flow and for convenience we refer to it as  \lq\lq QK(q=1.5)". The  fastest growing mode of the MagnetoRotational Instability  has a growth rate $|\omega_{max}| \propto |q\Omega|$, therefore the value of $q=1.9$ was selected to take advantage of this.  The $q=1.9$ flow is divided into the three named profiles \lq\lq QK(Ekman)", \lq\lq QK(Split)" and \lq\lq QK(q=1.9)" to reflect the choice of end ring speeds. \lq\lq Ekman" refers to the boundary layer that we expected to significantly perturb the ideal profile, as was seen in a prototype experiment with $h/(r_2-r_1) = 1$ ~\citep{Kageyama:2004}. \lq\lq Split" duplicates the end cap condition commonly used in Taylor-Couette experiments, such as with Taylor, Richard and Wendt (though Wendt had no upper cap, but a  free surface). In the \lq\lq QK(q=1.9)" profile the end rings are rotating at speeds intermediate to those of the cylinders. The particular choice of speeds were arrived at by  trial and error to best approximate the ideal Couette profile, Equation~\ref{eqn:ideal_couette}. Radial profiles of $v_\theta$ for the $q=1.9$ profiles will be presented in Section~\ref{sec:profiles}.

In addition to the quasi-Keplerian profiles, we also present measurements of $\Sigma_{r\theta}$ from three other anti-cyclonic flows: MS, CUS 1 and CUS 2.
They were chosen to lie at or above the Rayleigh centrifugal stability limit. MS refers to a configuration for which the end ring and inner cylinder speeds have been scaled up to the marginal stability line, $q=2$. For the cylinder radii we have chosen, the Rayleigh criterion is reached when the inner cylinder speed is $\Omega_1 = \Omega_2 r_2^2/r_1^2 = 8.27$, the CUS profiles are centrifugally unstable by this criterion. For CUS 1, the speed of the inner cylinder  and the end rings have been scaled by a constant from those of the QK(q=1.9) configuration by 15\% above the Rayleigh criterion. To make the flow even more unstable, the CUS 2 profile has had the outer end ring and outer cylinder speeds set to zero. Radial profiles of $\bar{v}_\theta$ for the CUS  cases were not measured. Measurements of transport in the turbulent centrifugally unstable cases are reported in \citet{Burin:2010}.

The flow profiles are established through an impulsive acceleration of the rotating components. Most often, the experiment was started from rest, though a transition from solid body to differential rotation was also common. 
The abrupt change in boundary speeds causes the fluid to spin up via Ekman pumping rather than viscous diffusion, though shear instabilities at the walls may contribute to spin-up at early times.
 The Ekman timescale governs the equilibration time, $\tau_{E} \sim \left(L^2/\nu\Omega_2 \right)^{1/2}$. When $\Omega_2=5.6$~rad/s, $\tau_E\sim 60$~s. We observed that the fluctuation levels reached their equilibrium values after approximately five viscous Ekman times, and so waited a minimum of 300~s before performing flow measurements~\citep{Schartman:2009}.
 
\begin{table}[hbtp]
   \centering
   \begin{tabular}{@{} l c c c c@{}} 
      \hline
      Profile name & $\Omega_1$&$\Omega_3$&$\Omega_4$&$\Omega_2$ \\
      \hline
      	QK(Ekman)	 & 7.50  & 1.00  & 1.00  & 1.00\\
	QK(q=1.9) 		 & 7.50 & 2.74 & 0.77 & 1.00 \\
	QK(Split) 		 & 7.50  & 7.50 & 1.00 &1.00\\
	QK(q=1.5) 		 & 5.91 & 2.48 & 1.09 & 1.00\\
	MS 		 & 8.27 & 3.02  & 0.83 & 1.00 \\
	CUS 1 	 & 9.43  & 3.43 & 0.94 &1.00\\
	CUS 2 	 & 9.43 & 3.43 & 0.00 & 0.00\\
	Solidbody & 1.00 & 1.00  & 1.00  & 1.00\\
	\hline
   \end{tabular}
   \caption[Anticyclonic fluid profiles]{ Component speeds used to produce  marginal and unstable flow profiles. Speeds are listed as a fraction of the Outer Cylinder speed, $\Omega_2$. MS is marginally stable, CUS is unstable by the Rayleigh criterion. } 
   \label{tab:speed_ratios}
\end{table}

\subsection{Diagnostics}

\subsubsection{Laser Doppler Velocimetry}

We use LDV for two types of measurement depending on the orientation of the diagnostic with respect to the apparatus, see Figure~\ref{fig:apparatus}. When the diagnostic is oriented to view the fluid through the outer cylinder wall, radial profiles of azimuthal velocity, $v_\theta$, are acquired. Viewing the fluid from beneath the vessel, we measure simultaneously $v_\theta$ and $v_r$ with a correlation window of $10 \mu s$ between them. Thus, all fluctuations are measured up to the frequency of $10^5$ Hz, which is comparable to the Doppler frequency of structures at the viscous scale in water or the size of the tracer particles used.
All flows are calibrated against solid body rotation. 

In this experiment, the primary sources of error in the LDV measurement are due to optical properties of the acrylic outer cylinder. 
The errors arise from the curvature of the cylinder, misalignment of
the probe head with respect to the vessel and random defects in the
optical path. The random defects arise  through stress-induced
variations in the  refractive index of the acrylic.  The variations
most likely induce spurious correlations by  altering the focal
lengths of the LDV beams~\citep{miles1996geometry}. The velocity
measurement is proportional to $\sin(d/2f)$, where $d$ is the beam
separation and $f$ is the focal length. Varying the focus for both
$v_r$ and $v_{\phi}$ measurements simultaneously will introduce a
simultaneous over or underestimation of the velocities.
The mean error is removed by calibration against solid body rotation.
However, the random defects broaden the sample distribution of the velocity measurement, which cannot be eliminated. The limiting precision of LDV is $v^\prime/v\approx 0.1\%$, whereas our precision was 1.5\%.
Details of the calibration process can be found in~\cite{Schartman:2008}.


When the LDV was setup to measure radial profiles of $v_\theta$, the beam probe head is oriented to view the fluid through the curved outer cylinder wall. 
The laser paths for the velocity component tangent to the azimuthal angle enter and exit the cylinder wall at non-normal incidence. 
As the probe is scanned radially the angles of incidence change, which alters both the radial location and magnitude of the $v_\theta$  measurement, see~\cite{Schartman:2008}. The location of the axial velocity measurement did not coincide with that of $v_\theta$ so $\Sigma_{\theta z}$ could not be measured. Moreover, the $v_z$ measurement location was often outside of the flow volume, so this velocity component was ignored. 

When the LDV diagnostic is  oriented axially to measure $v_\theta$ and $v_r$, the beam path passes through the cylinder bottom cap and outer ring. 
Attenuation of the laser intensity due to the four boundary crossings limits the data rate to order 1~Hz. 
For the previously reported measurements of the radial-azimuthal component of Reynolds stress \citep{Ji:2006,Burin:2010}, we required of order $10^3$ velocity samples per fluid profile. 
Because of this time demand, LDV was performed at only one radial location in this orientation. 
 The LDV probe head was oriented such that the beam bisector was approximately normal to the acrylic surfaces. 
 Comparison with the expected velocity value for solidbody rotation revealed an offset of about 3\% in the magnitude of the $v_\theta$. 

\subsubsection{Flow visualization}

Flow visualization of CU, QK(Ekman) and QK(q=1.9) profiles was performed using water with a 0.3\%  addition of Kalliroscope  AQ-1000 rheoscopic concentrate. The Kalliroscope flakes  align   with the local fluid  shear. The intensity and direction of  light scattered by the flakes  to make flow patterns visible~\citep{Savas:1985}. In laminar circular Couette flow the surface normal of the flakes align parallel to $\hat{r}$. Illuminated from below,  ideal (and stable) Couette flow should appear as a uniform color.   The flow over the outer ring was illuminated from below by a flashlight and imaged from the side by a CCD.

An attempt to image the flow over the inner ring using this method was not successful. The broad area of illumination combined with the low Kalliroscope concentration required to view the inner region of the flow could not produce enough contrast in the CCD to generate useful images.

\subsection{End caps effects: Ekman and Stewartson layers \label{sec:boundary_flow}}

All physically realizable Taylor-Couette experiments have a finite cylinder height, $h$. Rigid plates form the vertical boundaries, with some exceptions such as \citet{Wendt:1933} who used a free upper fluid surface. The non-slip boundary conditions at the end caps may introduce deviations from the ideal Couette profile, Equation~\ref{eqn:ideal_couette}. The influence of the boundaries is never completely negligible even for  aspect ratios in excess of 100~\citep{Taylor:1936, Tagg:1994}. Two possible effects of the boundaries are to produce Ekman~\citep{Greenspan:1968} or Stewartson layers \citep{Stewartson:1957}. 

\begin{figure}[htbp] 
   \centering
   \includegraphics[width=4in]{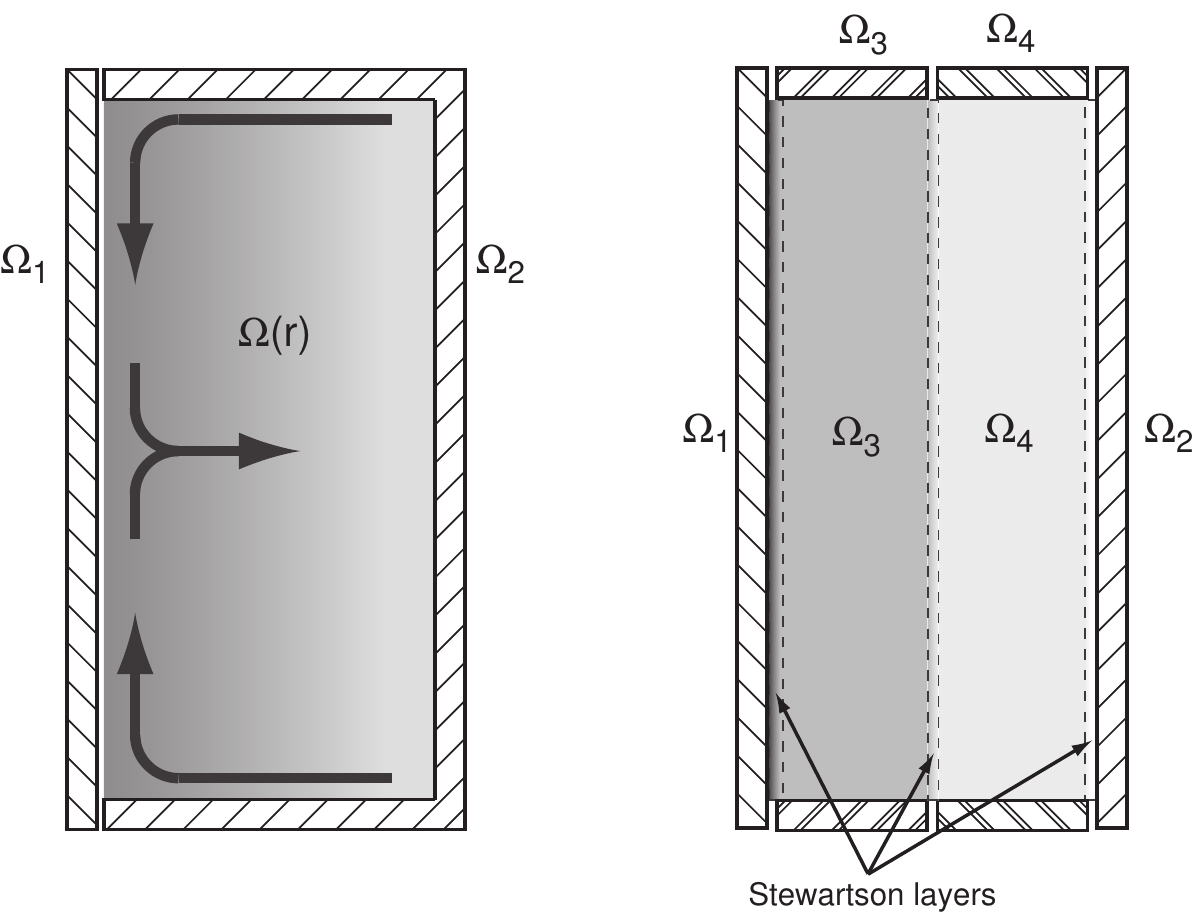} 
   \caption{End cap effects which were predicted to distort significantly  the mean profile of $v_\theta$ from the ideal Couette solution. Left: In the bulk flow the velocity profile is supported by the radial pressure gradient, $v^2_\theta(r)/r = \partial p/\partial r$. The pressure is approximately uniform vertically, but at the end caps the non-slip boundary condition requires $v_\theta = r\Omega_2$. This imbalance at the end cap drives the radial  Ekman circulation, when $\Omega_2 < \Omega_1 $ the flow is directed inward. When the Ekman flow reaches the inner cylinder it transitions to an axial flow. In a small aspect-ratio device such as the Princeton MRI Experiment ~\citep{Kageyama:2004} predict that the Ekman circulations  would meet at the midplane and form a radial jet. 
Right: With  independent end rings, the strength of the Ekman pumping can be reduced by minimizing the velocity difference between the rings and the bulk flow. The simulations of \citet{Kageyama:2004} showed that two rings provided an effective trade off between mechanical complexity and reduced Ekman pumping.
A simulation of the two-ring design by \citet{Hollerbach:2004} predicted the formation of Stewartson layers at the boundaries between the components. The ideal Couette profile would then be replaced by two annuli in solid body rotation with the end rings. }
   \label{fig:boundary_effects}
\end{figure}

In the bulk flow pressure balances the centrifugal force of the velocity profile, $\partial P/\partial r = \rho r\Omega^2_{bulk}(r)$. The pressure is approximately constant vertically, while the velocity in the viscous layer attached to an end cap is rotating at the end cap speed, $\Omega_{cap}$. At the interface between the bulk flow and the viscous layer, a radial Ekman flow is generated to balance the discontinuity of force, see Figure~\ref{fig:boundary_effects}. If $\Omega_{cap} < \Omega_{bulk}$, the Ekman flow is directed inward radially. When the layer reaches the inner cylinder, it transitions to an axial flow and is accelerated as it spirals along the cylinder.
In large aspect ratio devices, the Ekman layer may detach from the inner cylinder before reaching the midplane \citep{Coles:1965}. In small aspect ratio devices such as our prototype~\citep{Kageyama:2004} we observed that the Ekman layers from each end cap propagated to the cylinder midplane. The flows at the juncture then detached from the inner cylinder in a radial jet. The advection of angular momentum by the Ekman layers produced a mean radial profile of $v_\theta$ which was centrifugally unstable near  the inner cylinder, and flattened over the larger radii.   The cylinders of the current apparatus are only a factor of two taller than the prototype, so we expected a similar effect occur when both end rings co-rotate with the outer cylinder.

Stewartson layers arise from the tendency of a rotating flow to be uniform along the axis of rotation, at least on timescales long compared to an Ekman time, $\tau_E$. 
Simulations  by 
\citet{Hollerbach:2004} of the Princeton MRI Experiment predicted the formation of Stewartson layers at the boundaries of the end rings, see Figure~\ref{fig:boundary_effects}. The layers are predicted to  separate the bulk  fluid in to two annuli. The inner annulus would be in solid body rotation with the inner end rings, and the outer annulus with the outer end rings. The narrow shear layers between the two annuli, and between the annuli and cylinders would be unstable~\citep{Hide:1967,Fruh:1999}.

Two experiments using LDV were performed to determine the extent to which Ekman or Stewartson layers cause the bulk flow to deviate from the ideal Couette profile. The first experiment was to compare the radial profiles of $v_\theta$ using end ring speeds which would maximize or minimize the strength of the Ekman circulation. The second experiment was a detailed vertical scan of $v_\theta$ near the ring gap for the profile which minimizes the Ekman circulation. The results of these experiments are presented in Section~\ref{sec:profiles}. 



\section{Results}

\subsection{Topography of the QK(q=1.9) profile}

\begin{figure}[htbp] 
   \centering
   \includegraphics[width=\textwidth]{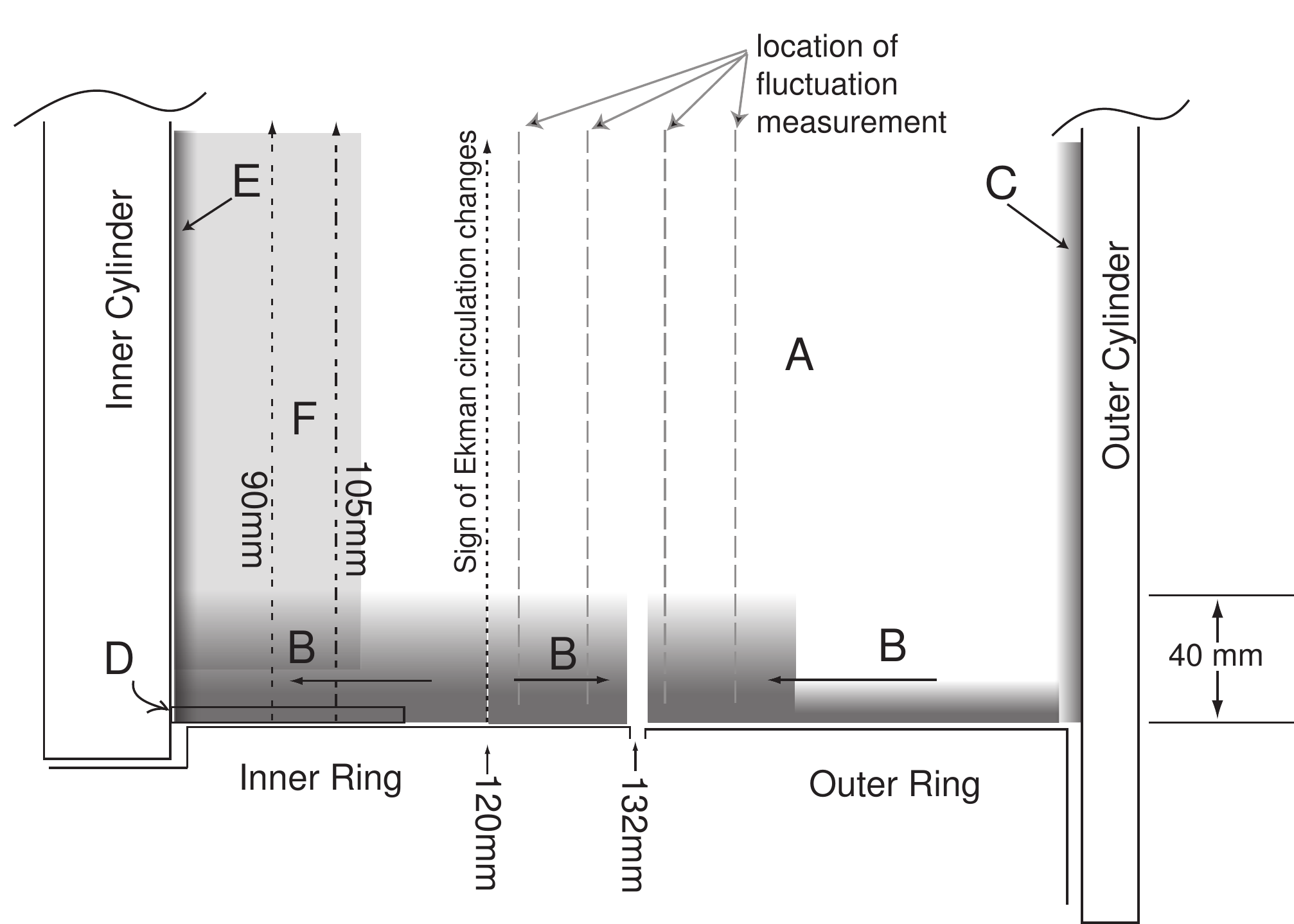} 
   \caption[Regions of flow for MRI operation]{
   Regions of flow within cylinder gap MRI operation, in this case the outer ring is rotating slower than the bulk flow and the inner ring corotates with the bulk flow at $r=120$~mm. The lower half of the apparatus is shown. A) The Taylor-Proudman theorem holds in the bulk flow, which closely approximates the ideal Couette profile. B) Fluctuations generated by the vertical boundaries extend approximately 40~mm into the flow. The sign of radial circulation within the "boundary zones"  are indicated by the horizontal arrows. 
  At $r=120$~mm, the boundary corotates with the bulk flow. The radial velocity of the Ekman circulation changes sign at this radius as the ring is removing (adding) angular momentum to the bulk at smaller (larger) radii. The Ekman circulations penetrate into the bulk flow, see Figure~\ref{fig:gap_scan}.
   Near the ring gap a vertical scan of $v_\theta$ fluctuations indicates that the discontinuity in velocity produced at the ring gap does not propagate in to the bulk flow.
 C) Centrifugally stable boundary which transitions from the bulk flow to the outer cylinder.  (This region C may be centrifugally unstable when the cylinder speeds exceeds the Rayleigh criterion as in the flow CUS 1 listed in Table 2. This leads to a lower $q$ than 2 in the region A, and may explain the small measured Reynolds stress there as shown in Fig.\ref{fig:beta_plot}). D) Ekman layer detaches from walls with higher fluctuations visible in the 5~mm radial scan, see Figure~\ref{fig:vert_scan}. The maximum fluctuation amplitude due to the detachment is approximately 37\% and occurs at $r\approx 105$~mm~\citep{Schartman:2008}. E) Measurement of $q$ indicates a centrifugally-unstable region exists between the inner cylinder and $r_1 \approx 90~\mathrm{mm}$. F) The unstable flow relaminarizes by $r=100$~mm. 
   }
   \label{fig:flow_regions}
\end{figure}

Before presenting the details of the experiments, the topography of a flow generated in the QK(q=1.9) configuration is diagrammed in Figure~\ref{fig:flow_regions}. The QK(q=1.9) profile is displayed because it demonstrates the greatest diversity of behavior in the apparatus. The profiles generated by a different choice of end ring speeds feature one or more of the behaviors of the QK(q=1.9) profile.
 The profile consists of a bulk flow in which the Taylor-Proudman theorem holds, and is bounded at the cylinder walls  by thin shear layers. The layer at the inner cylinder is anti-cyclonic and unstable by Rayleigh's criterion. The layer at the outer cylinder is cyclonic, and therefore stable. Ekman layers are present on the top and bottom end rings. One radius of the inner ring corotates with the bulk flow. On either side of this radius the Ekman flows change sign and the flows are able to penetrate deeply into the bulk flow. No radius of the outer ring corotates with the bulk flow, confluence of Ekman circulation does not occur.


The Kalliroscope images of Figure~\ref{fig:kalliroscope}(C) indicate the most obvious characteristic of the flow: a bulk profile in which the Taylor-Proudman theorem holds, is bounded above and below by a turbulent layer.  The measurement of $\Sigma_{r\theta}$ which extended from $z=35$~mm  up to $z=90$~mm shows no evidence of turbulent radial transport. Mechanical limits of the LDV diagnostic prevented measurements closer than 35~mm to the boundary. From Figure~\ref{fig:vert_scan} the region in which the Taylor-Proudman theorem holds extends to at least 5~mm from the end rings. We also  know from  Figure~\ref{fig:gap_scan} that the  fluctuation levels of $v_\theta$ are increasing  as the end ring is approached. Both of these imply the existence of a vertical Reynolds stress $\Sigma_{\theta z}$. Our ability to place a bound on $\Sigma_{\theta z}$ is limited by the uncertainty in the magnitude of $\bar{v_r} = 0.18\pm0.17$mm/s~\citep{Schartman:2008}. 


\subsection{Flow visualization}
Images taken with the Kalliroscope technique are shown in Figure~\ref{fig:kalliroscope}. The three flows shown are a CUS~2 centrifugally unstable flow with  the outer cylinder at rest, QK(Ekman) flow with the end rings co-rotating with the outer cylinder, and a QK(q=1.9) profile with both rings in differential rotation. The Reynolds numbers of the profiles are all approximately $5\times 10^5$.

In the centrifugally unstable case, the turbulence is fully developed. There are no visible large scale flow patterns. As discussed in \citet{Burin:2010} this lack of structure  is consistent with other experiments of  supercritical flow at large Reynolds numbers~\citep{Lewis:1999}.
The QK(Ekman) case has less structure than the QK(q=1.9) flow, though we will show in Section~\ref{sec:reynolds_stress} that a larger Reynolds stress is present in the QK(Ekman) flow.  In the QK(q=1.9) case the flow appears turbulent within about 40~mm of the end ring boundary, but in the bulk of the flow the uniform horizontal streaks indicate a quiescent flow. Note also that only in the QK(q=1.9) case does enough light reach the top ring to discern the upper turbulent band. This demonstrates that the flakes are most aligned with the rotation axis, consistent with the hypothesis that the velocity shear is radial.

\subsection{Radial profiles of quasi-Keplerian flows\label{sec:profiles}}

A comparison of three quasi-Keplerian flows is plotted in Figure~\ref{fig:v_SEM}. The speeds of the inner and outer cylinders are 42.0 and 5.6~rad/s, respectively. The Reynolds number is $Re=5.5\times10^5$ and the  nominal value for $q=-(\bar{r}/\bar{\Omega})\Delta\Omega/\Delta r =  1.9$.  The end rings rotate in the QK(Split), QK(Ekman), or QK(q=1.9) configurations.

Radial profiles of $V_\theta$ at several axial heights are plotted in Figure~\ref{fig:vert_scan} for the QK(q=1.9) configuration. The axial positions vary from 5~mm above the lower rings to the midplane at 140~mm.  The plotted profile is at $Re=3\times10^5$ with $\Omega_1 = 21.0, \Omega_3= 7.6, \Omega_4= 2.1, \Omega_1 = 2.7~\mathrm{rad/s}$. The thickness of the viscous Ekman layer is $\delta_{Ekman} = \sqrt{\nu/\Omega}\sim4~\mathrm{mm}$. The profiles indicate that the Taylor-Proudman theorem holds in the bulk of the flow, but does not all the way to the end rings. This is especially true for the outer ring which is rotating slower than the outer cylinder. The deviation from the mean profile near the inner cylinder in the 5~mm data is due to an intermittent presence of the boundary layer as it is ejected from the inner cylinder-inner ring transition, see~\cite{Schartman:2008}. The plotted profile is stable by the Rayleigh criterion but two vertical shear layers are present at the cylinder walls.

A detailed vertical scan was performed near the ring gap to look for evidence of the Stewartson layer predicted by \cite{Hollerbach:2004}. The results of that scan are plotted in Figure~\ref{fig:gap_scan}. The scan was performed at four radii: $r$~=~123, 129, 134 and 140~mm which are shown in  Figure~\ref{fig:flow_regions} as the vertical dashed lines. For the radii 129, 134 and 140~mm the fluctuations levels, $\sigma_{v_\theta}/\langle v_\theta \rangle$, are comparable to the solid body fluctuation amplitude which is plotted as the vertical dotted lines. No evidence of a Stewartson layer is present near the ring gap at 132~mm. 
 For the points at $r$~=123~mm, the fluctuation amplitude does not converge to the solid body value but remains slightly elevated through to the experiment midplane. The reason for this is that a Stewartson layer does form in the vicinity of the radius at which the end ring co-rotates with the bulk profile, $\Omega_3 = \Omega(r=123~\mathrm{mm})$. A single radial scan using a QK(Ekman) profile is also plotted, marked by the squares. In the QK(Ekman) configuration the end caps rotate with the outer cylinder and because of the absence of a stationary radius in the bulk flow, the fluctuation level is comparable to the sold body level.
The fluctuation levels of the Stewartson layer at 123~mm are comparable to fluctuation levels over the outer ring in CUS flows in which the outer cylinder is rotating~\citep{Schartman:2008}.

To perform the measurement of $\beta$ given in Equation~\ref{eqn:beta_measure} we need the $\Sigma_{r\theta}$ component of the Reynolds stress and a value for the local exponent of angular velocity, $q$. Correlated two-component LDV provides the measurement of $\Sigma_{r\theta}$. We determine $q$ from the radial profiles of $v_\theta$ plotted in Figure~\ref{fig:v_SEM}. Because of the proximity of the radial boundaries it is tempting to compute an average value based on the cylinder speeds: $q\approx-(\bar{r}/\bar{\Omega})\Delta\Omega/\Delta r$. Using this formulation, the three profiles in the figure have the same value for $q$. A cursory inspection shows that the QK(Split), QK(Ekman) and QK(q=1.5) profiles do not share the same gradient at $r=180$~mm. 

Instead we compute $q$ based on a cubic-spline interpolation of $v_{\theta}$.  Determination of $q$ was also done using a finite difference scheme, see~\cite{Schartman:2008} with similar results. The interpolation of the profile is plotted as the dotted lines in Figure~\ref{fig:v_SEM}. Radial profiles of $q$ for the QK(Split), QK(Ekman) and QK(q=1.9) configurations are plotted in Fig.~\ref{fig:q_SEM_comp}. We note that the $q$ value for the QK(q=1.9) configuration is actually larger than 2 near the outer cylinder but without apparent local centrifugal instabilities.

\subsection{Reynolds stress\label{sec:reynolds_stress}}

Errors in the measurement of $\beta$ are dominated by random and systematic errors in $\langle v_r v_{\theta} \rangle$. We reduced the random error  by acquiring of order $10^3$ samples for each measurement.  The systematic error is removed by subtracting the correlation level measured in a solid body profile. The sources of the systematic error include any defects in the acrylic vessel and rings along the laser optical path and any residual inherent correlations between the two components of LDV systems as discussed in
Sec. 3.3.1. Example data and their comparisons can be found in Fig.~3 from \citet{Ji:2006}.
The transport levels for linearly-stable profiles are plotted in Figure~\ref{fig:beta_plot}. The measurements are performed at $r=$179mm and $z=62$mm or $z$=74mm, but the results are robust for other locations, see Fig.~3 from \citet{Ji:2006} for the $z$-dependence. The optimized QK(q=1.9) and QK(q=1.5) configurations for $Re > 10^6$ are not consistent with Richard and Zahn's proposed  transport level.  Averaging the results for the optimized configurations yields $\beta =(1.13 \pm 1.15)\times10^{-6} $ and $\beta < 3.4\times10^{-6}$ at 2 standard deviations. Because negative values for $\beta$ would indicate inward transport of angular momentum (which we believe to be unphysical, as it would intensify rather than extract energy from the
shear), 2 standard deviations yields 98\% confidence. The slight improvement of the transport limit over the previously published result~\citep{Ji:2006} is due to the use of the local measurement of $q$ and a correction for a small angular misalignment of the LDV diagnostic. 

Choices of the ring speeds other than the optimized values have profound effects, through the global Ekman circulation, on the mean flow profiles (Figure~\ref{fig:v_SEM}) as well as on the Reynolds stress, $\Sigma_{r\theta}$.
When the ring speeds are chosen to be equal to the outer cylinder [the QK(Ekman) configuration], 
$\Sigma_{r\theta}$ is raised significantly above the zero line as shown in Figure~\ref{fig:beta_plot}, 
even above the $\Sigma_{r\theta}$ value measured in the CUS 1 configuration which has a local $q$ below 2 in the bulk flow but with a centrifugally unstable layer near the outer cylinder (the region C in Fig.\ref{fig:flow_regions}).
We note that the $\Sigma_{r\theta}$ value in the QK(Ekman) configuration approaches the level proposed by \citet{Richard:1999}.
When the inner (outer) ring rotates at the inner (outer) cylinder speed [the QK(Split) configuration],
$\Sigma_{r\theta}$ jumps to much larger values. It is important to note that the QK(Split) configuration has stability only with respect to the two cylinder speeds but is unstable locally, i.e. at the ring gap. Given this level of transport, the observations by \cite{Richard:2001}, which uses the split configuration, are likely to be caused by the influence of the end rings extending throughout the fluid volume.
As a side note, the axial boundary conditions are also important for cyclonic flows briefly studied in the Princeton MRI
experiments. In the split configuration, radial profiles of these cyclonic flows deviated significantly from the ideal Couette profiles with large fluctuations~\citep{Burin2006}, and this subject is studied in detail by a followup experiment~\citep{Burin2011}.

If a subcritical transition occurred in the range $3\times10^5 < Re < 2\times 10^6$, what should we expect to see in $\Sigma_{r\theta}$? Suppose the transport follows the behavior of the normalized pressure drop, or \emph{friction factor}, plotted in the Moody diagram of pipe flow~\citep{Moody:1944}. A simplified moody diagram is reproduced in Figure~\ref{fig:moody} which displays only the laminar behavior and two turbulent cases. The turbulent curves are the smooth wall case and one rough wall case in which the scale height of the roughness is 5\% of the pipe diameter. A jump should occur at the transition from the laminar value to a value of order a few times the laminar transport level. If the flow is bounded by rough walls the friction factor should approach constant or logarithmic behavior when the viscous boundary layer becomes thinner than the roughness scale height. This formed the basis of our astrophysical extrapolation in \cite{Ji:2006}. However, our end caps are not rough nor is rough-wall behavior applicable to accretion disks. Instead, once the transition occurs the transport should again be a decreasing function of Reynolds number. If a transition occurred in the proposed range we should see, according to the Moody diagram, $\Sigma_{r\theta}$ jump to a factor of 2 to 4 above the laminar transport level $\beta_{visc}$ (where
$\beta_{visc} = \nu/\bar{r}^3|\partial\Omega/\partial r| = \nu/2b$ shown in Figure~\ref{fig:beta_plot}) and then fall off as $Re$ is further increased. The absence of such a spike in Figure~\ref{fig:beta_plot} argues strongly against the presence of a transition. 

The source of the higher levels of transport associated with the glycerol runs are hinted at by the time-averaged radial velocities~\citep{Schartman:2008}. 
Of the QK(q=1.9) configurations, only for these low Reynolds number experiments is $\bar{v}_r$  statistically distinguishable from zero. For example, at $Re = 2.2 \times10^4$ with glycerol $\bar v_r = -2.84\pm0.35$~mm/s, whereas for water $\bar v_r=0.18\pm0.17$~mm/s at $Re = 3.3 \times10^5$. The $v_r$ distributions for a glycerol run are compared to a water run and a solid body water run in Figure~\ref{fig:glycerol_pdf_vr}. At $Re = 2.2 \times10^4$, fluctuations are clearly present in the negative tail of the distribution which give rise to the non-zero value of $\bar{v}_r$. At the same speed but in water (so that $Re = 3.3 \times10^5$), the fluctuations are more symmetric about the mean and the mean radial velocity has fallen by a factor of two.  
We interpret this to mean that the residual unsteady secondary circulation penetrates deeper into the bulk flow at lower $Re$, consistent with the results from numerical study by \cite{Obabko:2008} which was performed at relatively low $Re$.



\section{Discussions and conclusions}

When operating in the QK(q=1.9) and QK(q=1.5) configurations with minimized Ekman effects we find no evidence of significantly enhanced angular momentum transport due to the $r-\theta$ component of the stress tensor, even at $Re \sim 10^6$. 
The absence of the measured transport efficiencies significantly above their laminar counterparts provides no evidence for a turbulent transition in our quasi-Keplerian flows in the range $10^4<Re<10^6$
since turbulence should enhance the transport to release the free energy it results from.
If a subcritical transition occurs at a greater Reynolds number, experiments with Poiseuille flow indicate that at the transition the amplitude of the dimensionless turbulent transport will be a factor of a few above the laminar viscous value which decreases as $Re^{-1}$. The tendency of the normalized turbulent
intensity to vary inversely with the critical Reynolds number has been confirmed for cyclonic flows, and also for anticyclonic flows to a lesser extent, by the numerical simulations by \cite{Lesur:2005}. 
We therefore conclude that either a subcritical transition does not occur, or, if a subcritical transition does occur, the associated radial transport of angular momentum is too small to directly support the hypothesis that subcritical hydrodynamic turbulence is responsible for accretion in astrophysical disks.
 
This picture conflicts with the observations of \cite{Richard:2001} in which flow imaging demonstrated a turbulent transition in quasi-Keplerian flow near $Re \sim 10^4$. Richard attributes the turbulence to SHI, but alternative hypotheses were not ruled out. In particular, our measurements in Split end-cap operation indicate that a detached shear layer is present above the inner ring. Such layers are known to have a non-axisymmetric linear instability \citep{Hollerbach:2003,Schaeffer:2005}. A plot of QK(Split) configuration at $Re=3\times10^5$ is reproduced in Figure~\ref{fig:split_5}. Also plotted are the velocities of the end rings, as dashed lines. A centrifugally unstable layer is present near the outer cylinder  therefore  the flow in the vicinity of 200~mm is  strongly turbulent. This observation is confirmed by $\Sigma_{r\theta}$ measurement for this flow: only in the CUS 2 case was a greater level of turbulent transport observed. The inflection point at $r\approx85$~mm has relaxed significantly from the solid body line of the ring which indicates that this region is the source of the instability. The precise nature of the instability cannot be inferred from this saturated flow. As noted, an inflection point is linearly unstable. But it is also possible that flow undergoes an axisymmetric transition because the fluid in solidbody rotation with the end ring has effectively increased the inner cylinder radius. 

There are possible limitations in applying the our results directly to astrophysical disks due to experimental geometry. Typical cold and dense astrophysical disks, such as protostellar disks, are geometrically thin, following the ordering $\bar r \sim \Delta r \gg h$, which contrasts with the scale ordering for our experimental apparatus, where all three scales are similar, a point that was made by one of our referees (P.-Y. Longaretti) as well as by \citet{Mukhopadhyay:2011b}. We now discuss the role of experimental geometry on our conclusions, namely with respect to aspect ratio ($h/\Delta r$) and normalized curvature ($\Delta r/\bar r$).

The dependence of the subcritical transition and turbulent transport on aspect ratio ($h/\Delta r$) is essentially unknown due to the absence of a first-principles theory. However, conceptually, in the absence of vertical friction, the vertical scale $h$ should not play a role in the transition $Re$ or in radial turbulent transport due to the instability's origin in radial shear at least when $h$ is larger than both $r$ and $\Delta r$, as it is in our experiment. In disks, $h$ (scale height of disk, or some fraction thereof due to stratification) may play more of a role in limiting the size of the largest eddies (or other structures) in the $r-z$ plane \citep{Dubrulle:2005}. How this limitation would ultimately affect angular momentum transport is unknown, though a reduction in transport may be implied. 

In experiments, the ratio $h/\Delta r$ may serve as a proxy for the strength of secondary flows due to viscous boundaries. These secondary flows may perturb the flow and trigger instability in subcritical flows - \citet{Richard:2001} and \citet{Paoletti:2011} (see below) probably being examples of this. These boundary-driven flows contribute to the radial momentum transport. Examples to the contrary, where secondary flows suppress instability and transport, are not known in high-$Re$ Taylor-Couette flow. Our end-cap controls, when optimized, have allowed us to reach relatively high $Re$ while maintaining a quiescent state. Non-optimized end-cap conditions (similar to those used by other expreriments with less controllable boundary conditions), on the other hand, produce fluctuations due to finite $h$. Thus our optimized velocity measurements, both mean and fluctuating, are suggestive of ideal flow.

Even if $h$ is insignificant, the role of curvature is not, and it remains uncertain as to which radial length scale is relevant. \cite{Richard:1999} showed that the relevant length scale in the cyclonic flows [based on \cite{Wendt:1933} and \cite{Taylor:1936}] changed from $\Delta r$ to $\bar r$ for $\Delta r/\bar r > 0.05$. An inferred quadratic scaling above this boundary yields $Re_c \sim 6 \times 10^5$ for the parameters of our experiment, a value which we have exceeded. Thus, even taking into account the effects of flow curvature, the lack of the evidence of turbulence in our results is significant. It is unclear though to what extent scalings based on cyclonic data may be applied to the anti-cyclonic quasi-Keplerian regime.    

Recent experimental work by \cite{Burin2011} in cyclonic flows illustrates that in experiments with large $\Delta r$, only a fraction of the flow becomes turbulent, a fraction that reflects the fact that the shear gradient in these experiments is localized near one of the cylinders. Outside of this region the flow remains laminar even well past $Re_c$. Thus the scaling of \cite{Richard:1999} appears to be an overestimate. In any case, this is not relevant to disks, where the balance between gravitational and centrifugal forces enforces a nearly constant ratio of shear to rotation. 

\begin{sidewaysfigure}[htbp] 
   \vspace{6in}
   \centering
   \includegraphics[width=\textwidth]{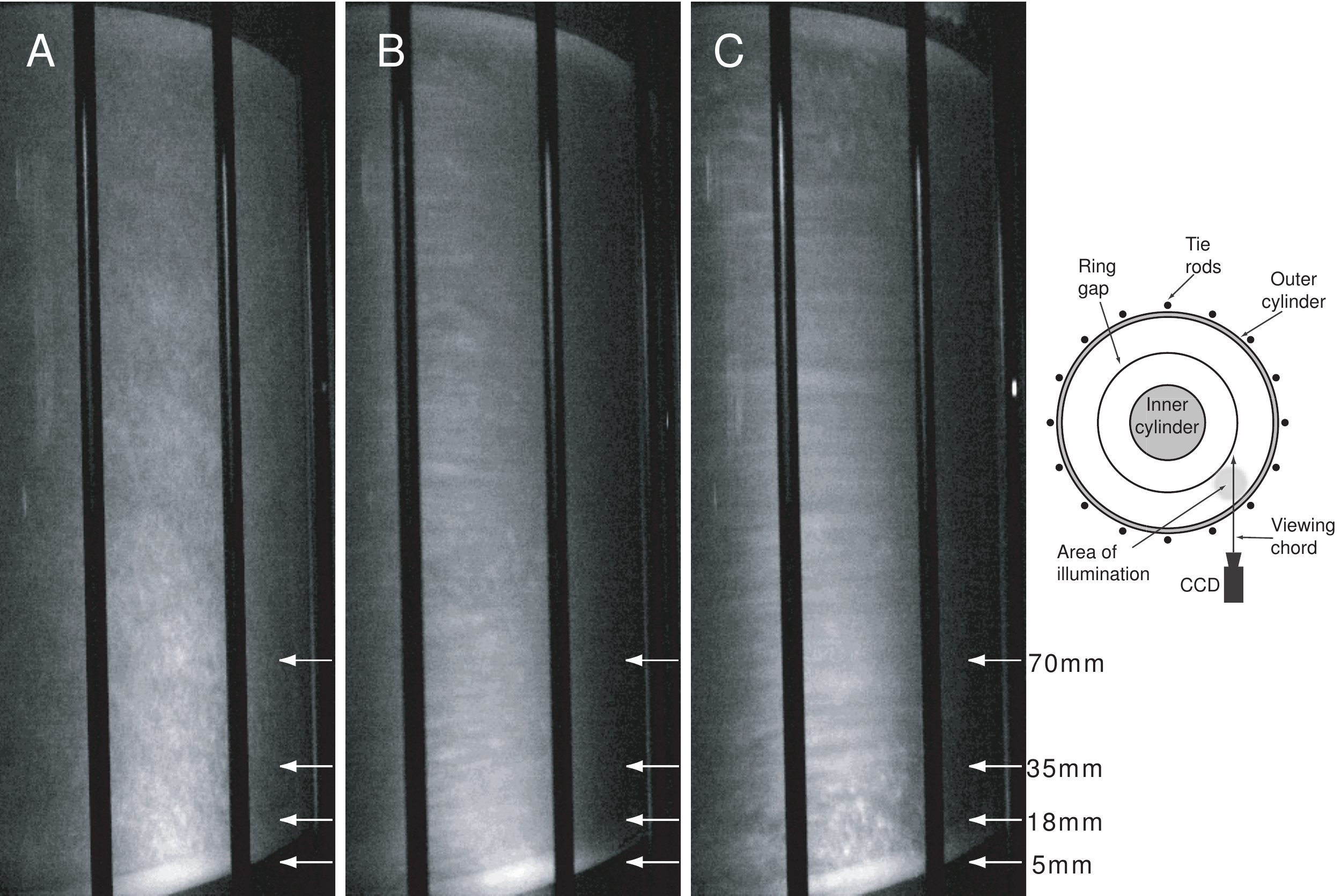}
   \caption{Kalliroscope image of three flows at $Re\approx5\times10^5$. A) centrifugally unstable flow with $[\Omega_1, \Omega_3,\Omega_4,\Omega_1]=[41.8, 0, 0, 0]~\mathrm{rad/s}$. B) A quasi-Keplerian flow using the Ekman configuration, QK(Ekman), $[\Omega_1, \Omega_3,\Omega_4,\Omega_1]=[41.8, 5.6, 5.6, 5.6]~\mathrm{rad/s}$. C) A quasi-Keplerian profile with end ring speeds optimized to produce the best approximation to the ideal Couette flow~\ref{eqn:ideal_couette}, QK(q=1.9), $[\Omega_1, \Omega_3, \Omega_4, \Omega_1]=[41.8, 15.3, 4.2, 5.6]~\mathrm{rad/s}$.} 
   \label{fig:kalliroscope}
\end{sidewaysfigure}

\begin{figure}[htbp] 
   \centering
   \includegraphics[width=4in]{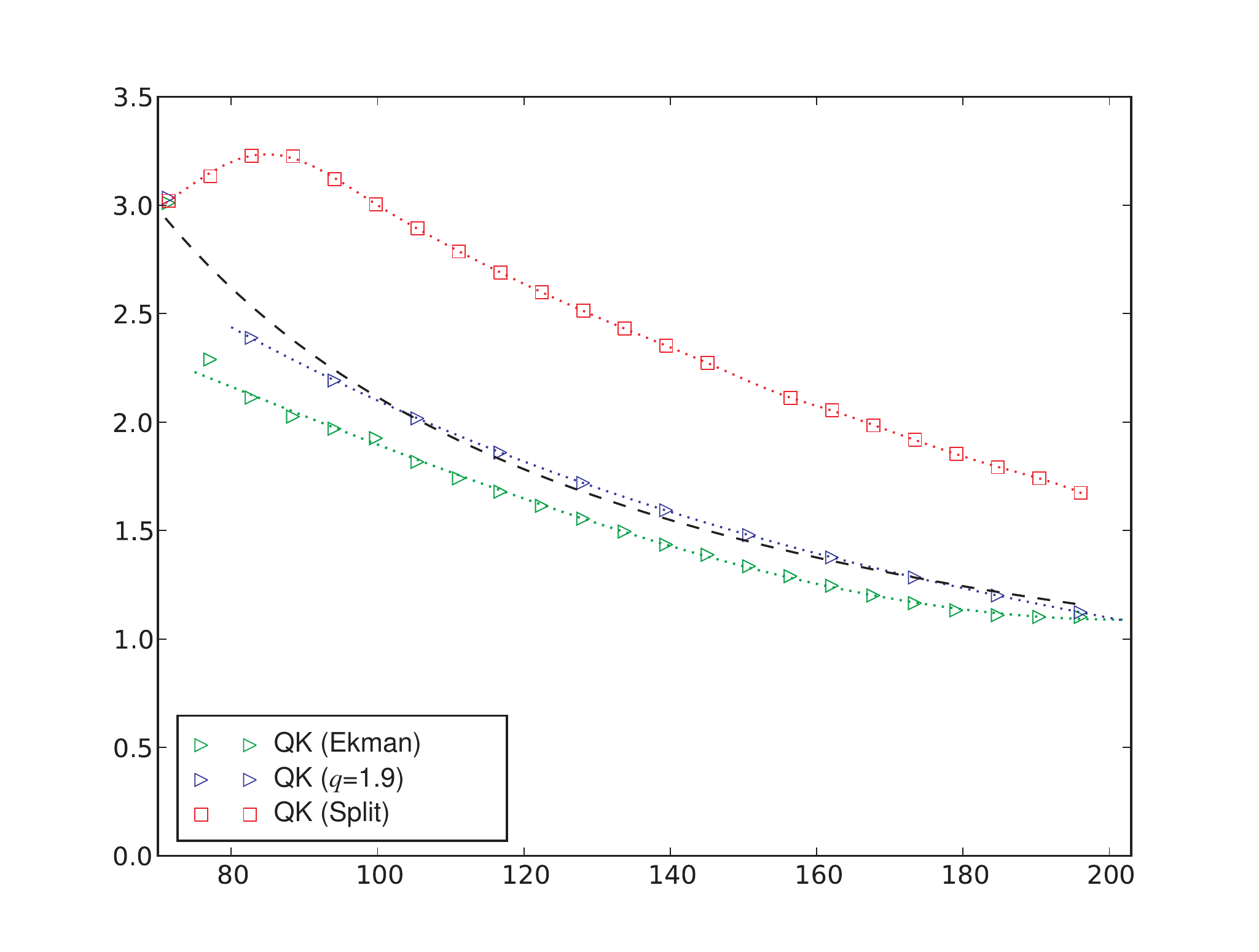} 
   \caption{$\langle v_\theta\rangle$ for three different centrifugally stable profiles at $Re=5\times10^5$. The dashed line is the ideal Couette profile based on the cylinder speeds, $\Omega_1 = 42.0, \Omega_2 = 5.6~\mathrm{rad/s}$. The dotted lines are cubic spline fits to the bulk flow. The innermost velocity measurement overlaps the inner cylinder and is therefore not included in the interpolation. Both the systematic and random errors in $\langle v_\theta\rangle$ are smaller than the marker size. }
   \label{fig:v_SEM}
\end{figure}

\begin{figure}[htbp] 
   \centering
   \includegraphics[width=3in]{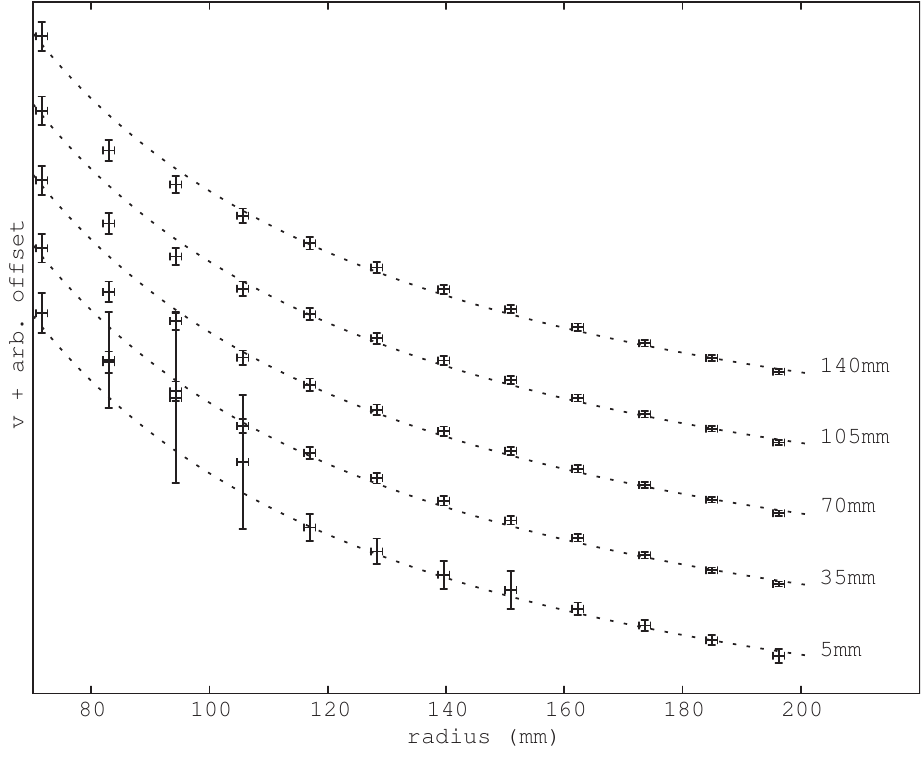} 
   \caption{Radial profiles of $v_{\theta}$ at several axial locations for an optimized flow (QK(q=1.9)) at $Re\approx3\times10^5$. To facilitate viewing the profiles have been offset. The increased fluctuation levels in the 5~mm scan near the inner cylinder are due to the disruption of the boundary layer circulation which occurs for differential rotation of the inner cylinder with respect to the inner ring. For the innermost radial point at 72~mm, the measurement volume of the diagnostic overlaps the inner cylinder and is therefore a measure of the surface speed of the cylinder rather than a flow velocity. }
   \label{fig:vert_scan}
\end{figure}

\begin{figure}[htbp] 
   \centering
   \includegraphics[width=4in]{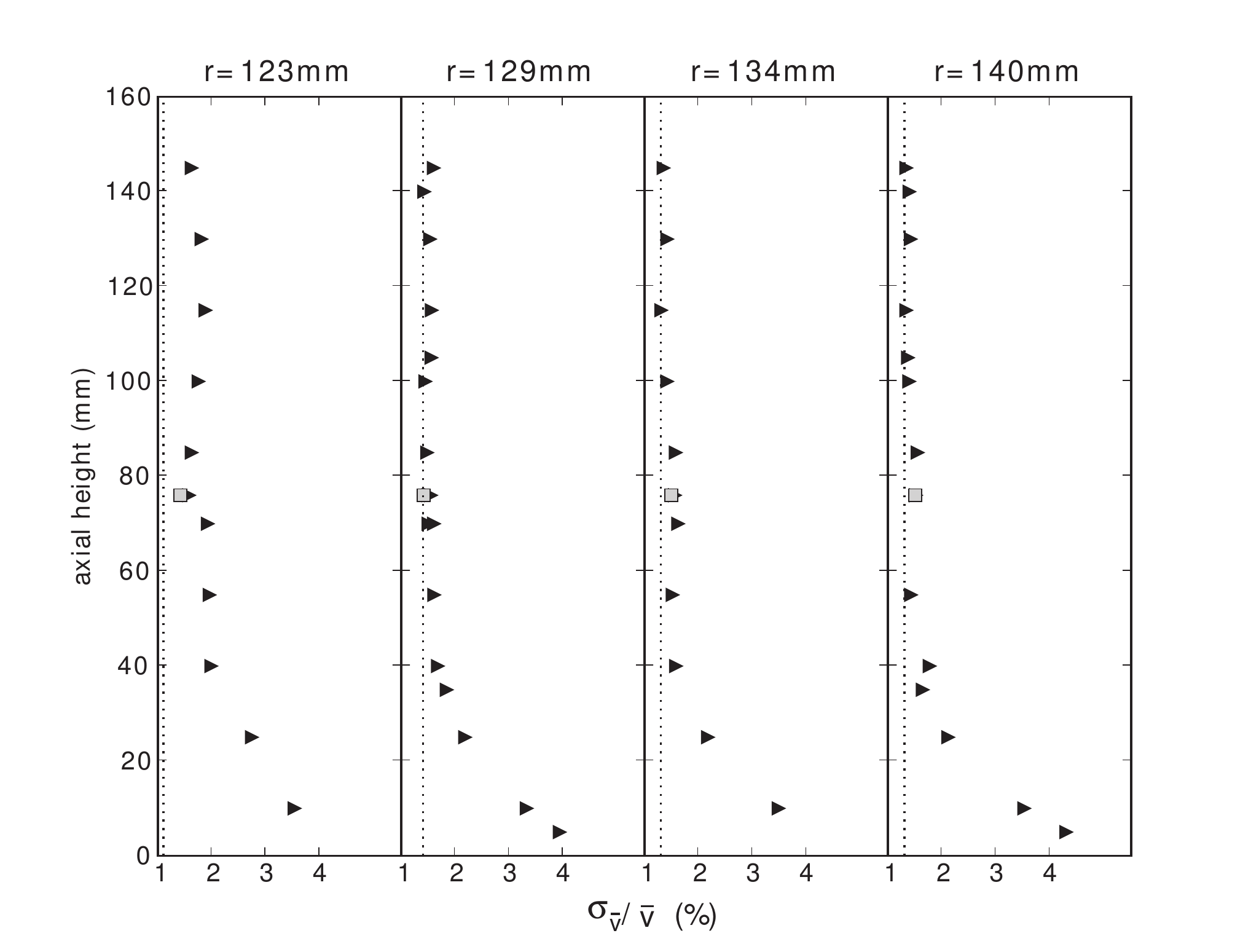} 
   \caption{Azimuthal velocity fluctuation levels, $\sigma_{v_\theta}/\langle v_\theta\rangle$, near the ring gap for the QK(q=1.9) profile at $Re=5\times10^5$ shown in Figure~\ref{fig:vert_scan}. The radial locations of the scans are diagrammed in Figure~\ref{fig:flow_regions}. Triangles are the QK(q=1.9) profile. Shown for comparison are the solid body levels, dotted lines and one axial location for a QK(Ekman) profile, squares.
   Note that in the r=123~mm panel, the solidbody line nearly overlaps the left plot border.}
   \label{fig:gap_scan}
\end{figure}

\begin{figure}[htbp] 
   \centering
   \includegraphics[width=3.75in]{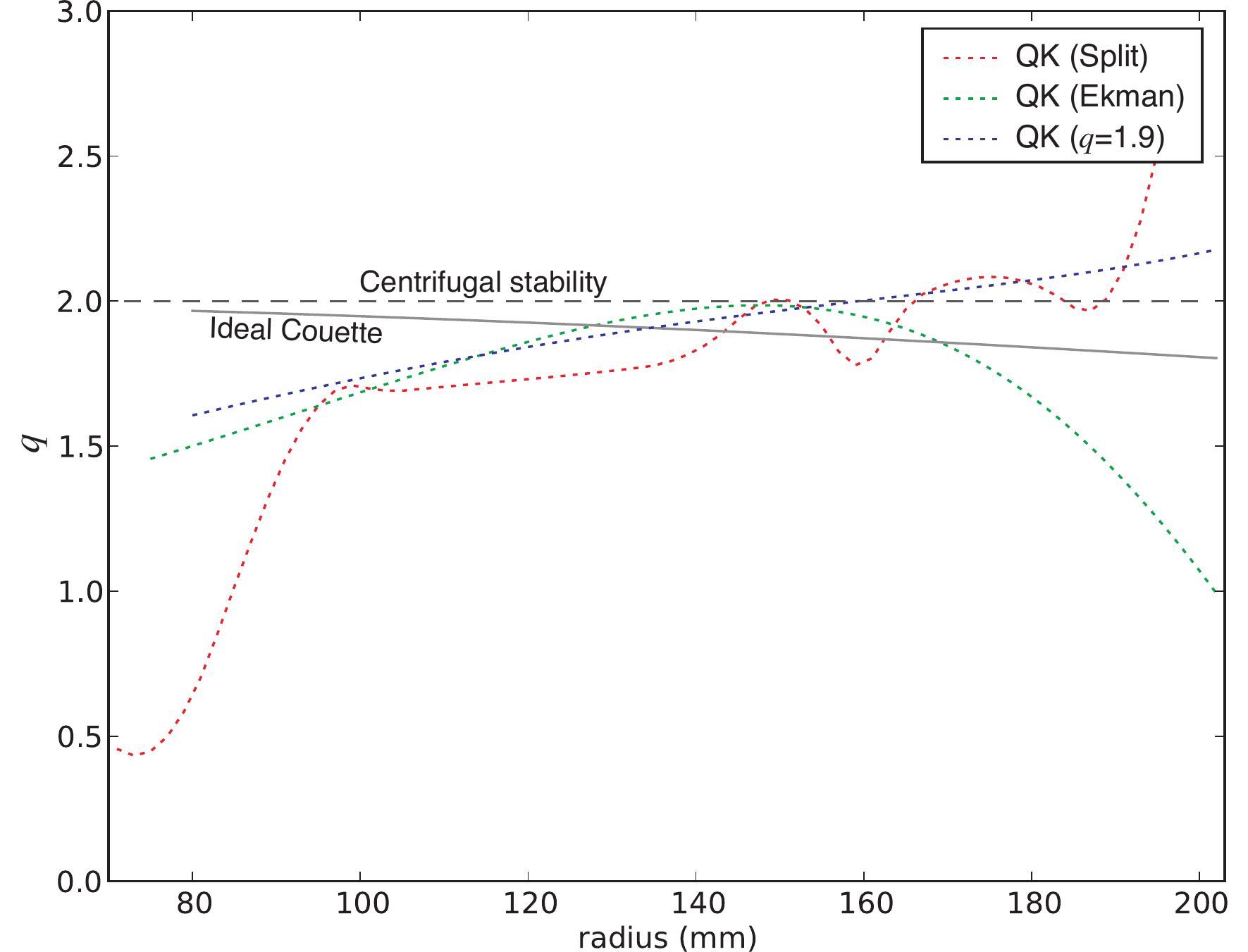} 
   \caption{$q$ profiles for centrifugally stable flows used in the calculation of $\beta$. The profiles are calculated using the spline fit presented in Figure~\ref{fig:v_SEM}.  \citet{Schartman:2008} used a  finite difference method to evaluate $q$ producing comparable results.   $q$ profiles are also plotted  for ideal Couette fow, and marginal stability by the Rayleigh criterion $q=2$.  The stability boundary derived by \citet{Rayleigh:1916} and \citet{Taylor:1923} do not account for the  presence of vertical boundaries. The oscillation in the QK(Split) case which occurs between 140 and 180~mm is due to the measurement of $v_\theta$ beginning before the flow had fully equilibrated.}
   \label{fig:q_SEM_comp}
\end{figure}

One might ask a subtle but relevant question, as raised by an anonymous referee, on whether our chosen axial boundary conditions for the ideal Couette profile 
might limit the transition or turbulent transport from occurring in the bulk flow. This question is motivated by the fact that
some well-known subcritical transitions in non-rotating flows are accompanied by large changes in the mean flow; see a recent study in a plane Couette flow by \cite{Krug:2012}.
While it is not possible to completely rule out such a possibility since the transition mechanism is not understood, but it seems unlikely.
The axial boundaries can influence the bulk flow either through global Ekman circulation 
or through local vertical vortex tubes in the spirit of Taylor-Proudman theorem. The former process
is slow on a timescale $\propto Re^{-1/2}$ as shown previously \citep{Kageyama:2004}, and 
it is unlikely to be dynamically important for the transition in the bulk flow. 
The latter process requires more rapid global rotation \citep{Spence:2012} 
or a strong axial magnetic field in an electrically conducting fluid \citep{Roach:2012}; none of them applies here.
In fact, the enhanced fluctuations observed near the axial boundaries [Figure~\ref{fig:kalliroscope}(C) and Figure~\ref{fig:gap_scan}] 
are indicative of the existence of
a turbulent boundary where Taylor-Proundman theorem should not apply.

\begin{figure}[htbp] 
   \centering
   \includegraphics[width=4in]{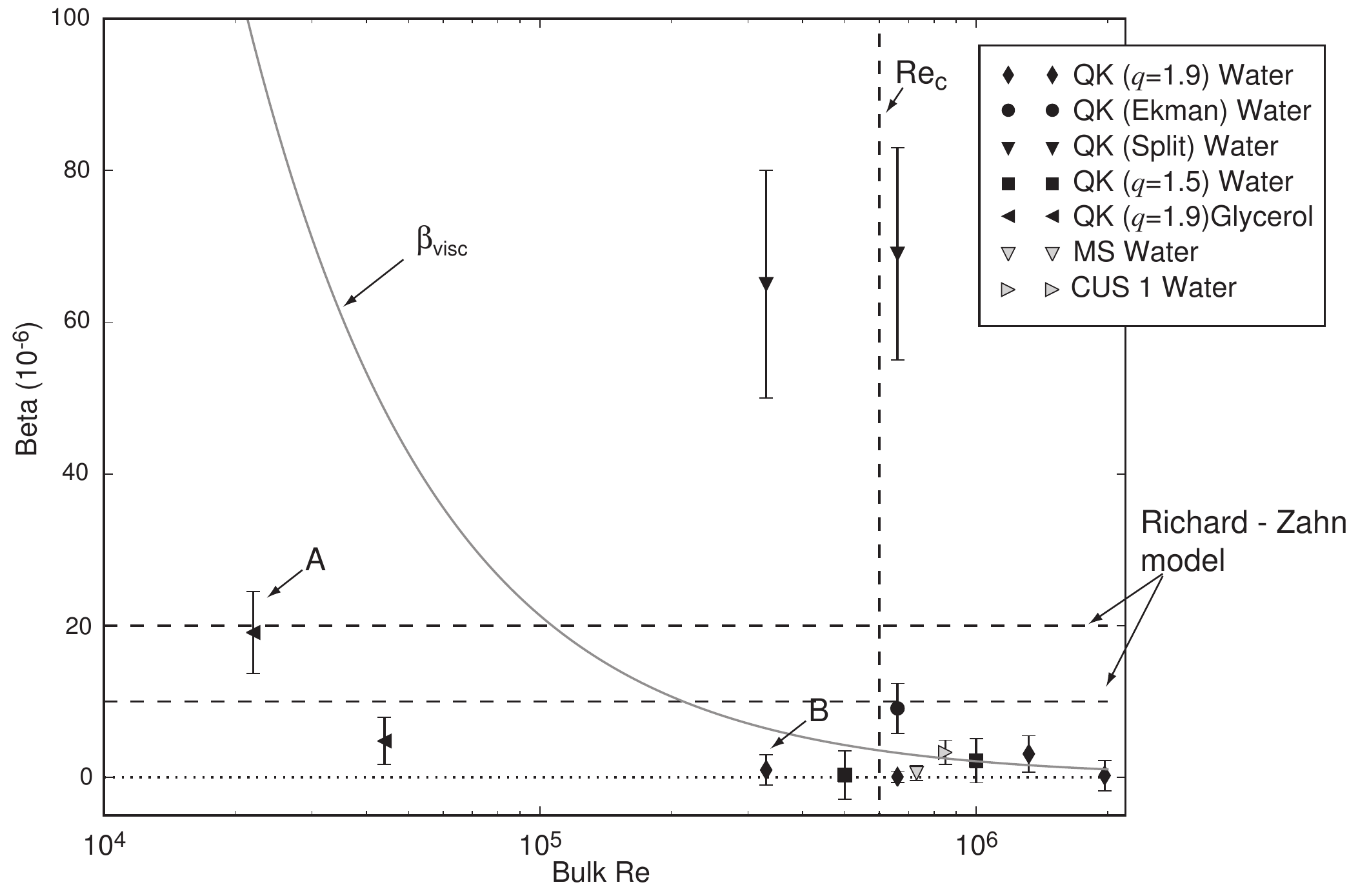} 
   \caption{Measurement of $\beta$ for quasi-Keplerian profiles at $r=179$mm and $z=62$mm or 74mm. Horizontal dashed lines is the predicted range of $\beta$ from Richard and Zahn. The vertical dashed line is the estimate of the transition Reynolds number predicted by Equation:~\ref{eqn:Re_c}. 
   }
   \label{fig:beta_plot}
\end{figure}

\begin{figure}[htbp] 
   \centering
   \includegraphics[width=4in]{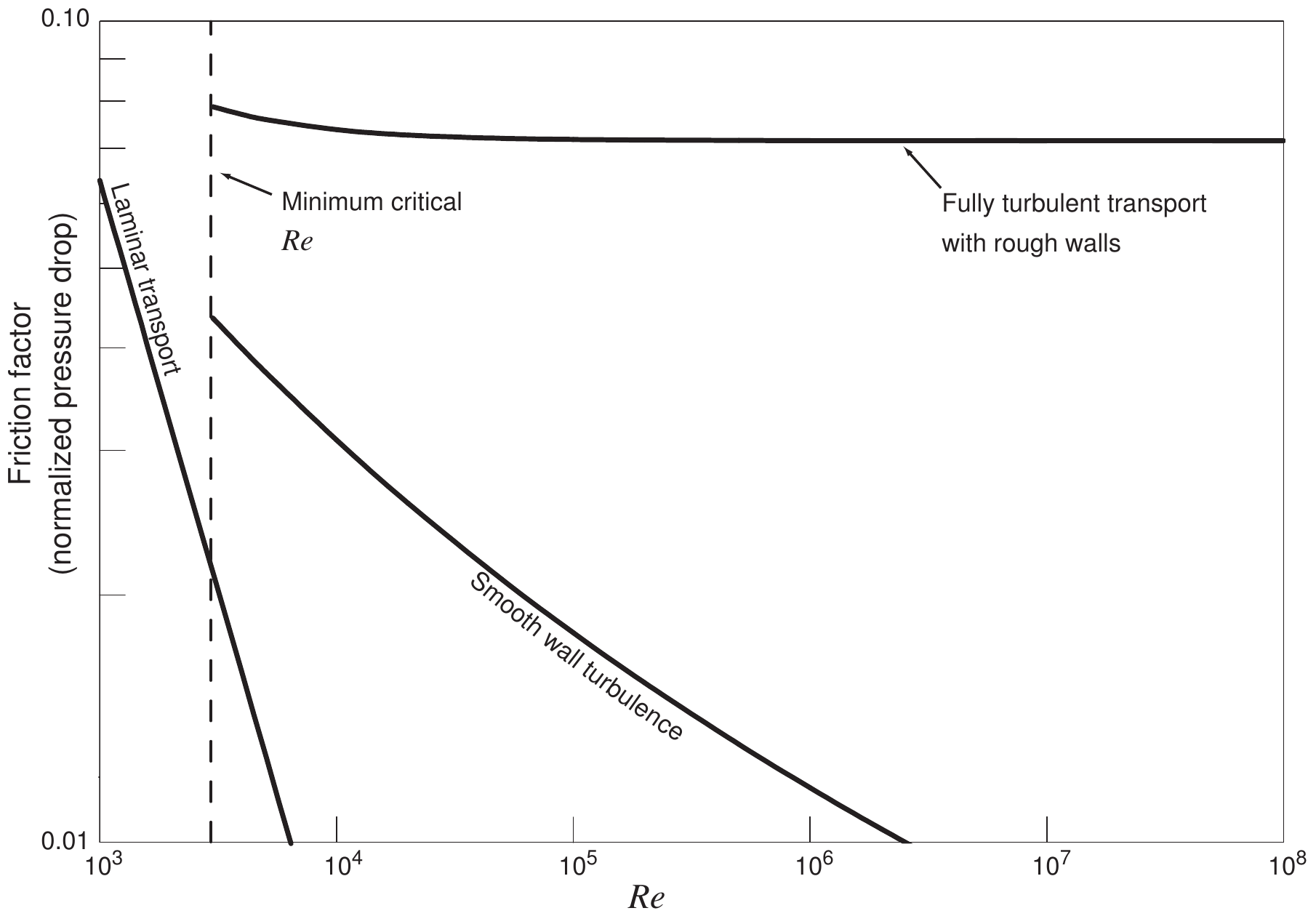} 
   \caption{A simplified Moody diagram for circular pipe flow. Only three transport curves are shown:  the laminar case, turbulence in a pipe with smooth walls, and one choice of rough wall turbulence. For the rough wall curve, the scale height of the roughness is 5\% of the pipe diameter. Below the minimum critical Reynolds number, a perturbed flow will always return to the laminar state. A transition to the turbulent state is marked by a discontinuity in the transport level. The minimum magnitude of the discontinuity in this case is approximately 2.}
   
   \label{fig:moody}
\end{figure}

\begin{figure}[htbp] 
   \centering
   \includegraphics[width=3in]{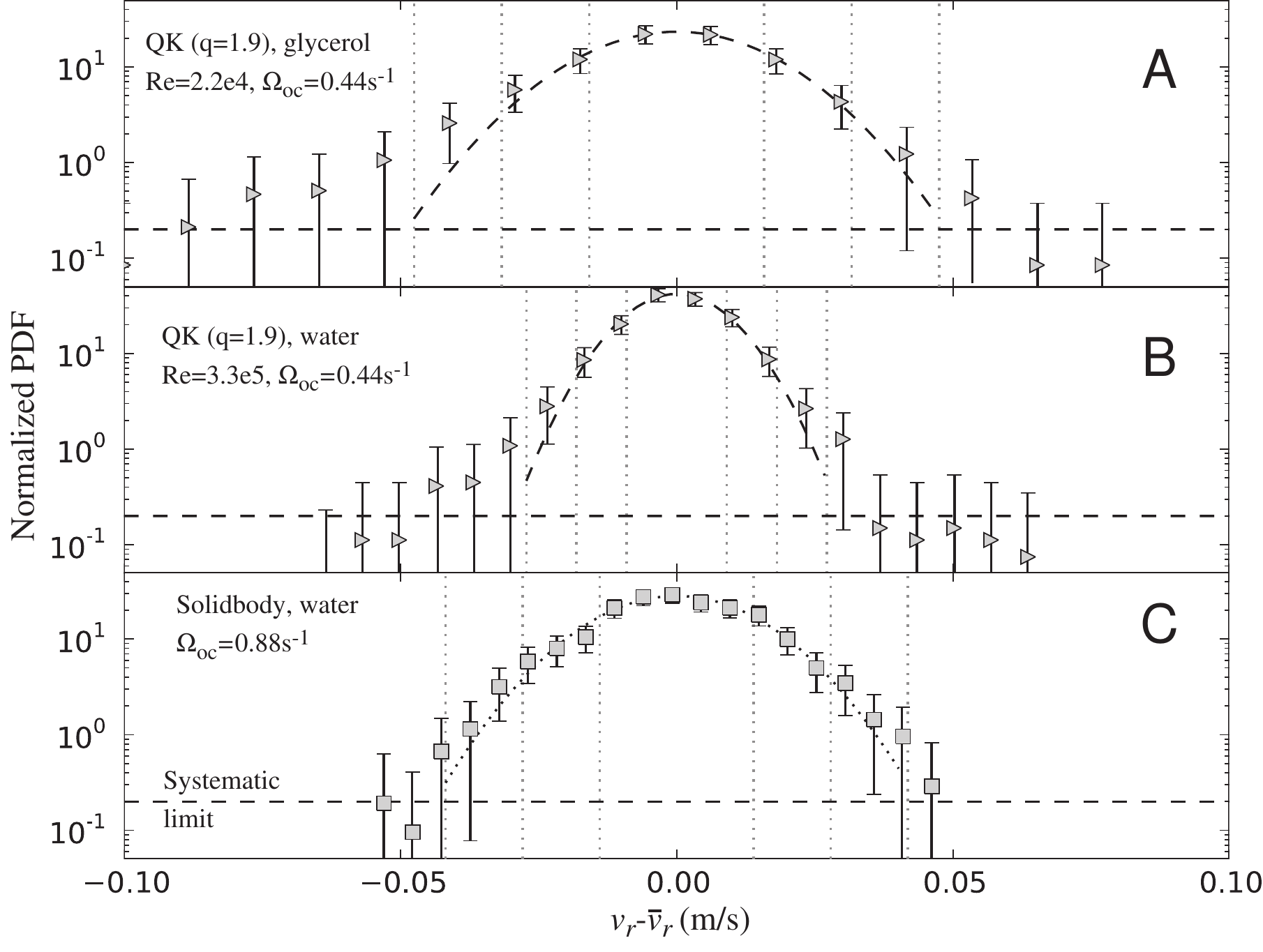}
   \caption{Probability densities of $v_r$ for QK(q=1.9) configuration at $Re\approx 2.2\times10^4$ using water-glycerol mixture (top panel, labelled {\bf A} in Figure~\ref{fig:beta_plot}), QK(q=1.9) configuration at $Re\approx 3.3 \times10^5$ using water (middle panel, labelled {\bf B} in Figure~\ref{fig:beta_plot}), and solid-body rotation (bottom). Apparently larger standard deviations in the solid body flow (C) than those in the QK flow in water (B) are due to the larger $\Omega_{OC}$ in the flow (C). The curves are Gaussian fits to the data, vertical lines are first, second and third standard deviations from their means $v_r =\bar v_r$. The data were acquired at $z=75$~mm. Note that non-Guassian tails at the level of $\approx0.2$ are due to acrylic optical defects and not genuine characteristics of the flows. }
   \label{fig:glycerol_pdf_vr}
\end{figure}

\begin{figure}[htbp] 
   \centering
   \includegraphics[width=3in]{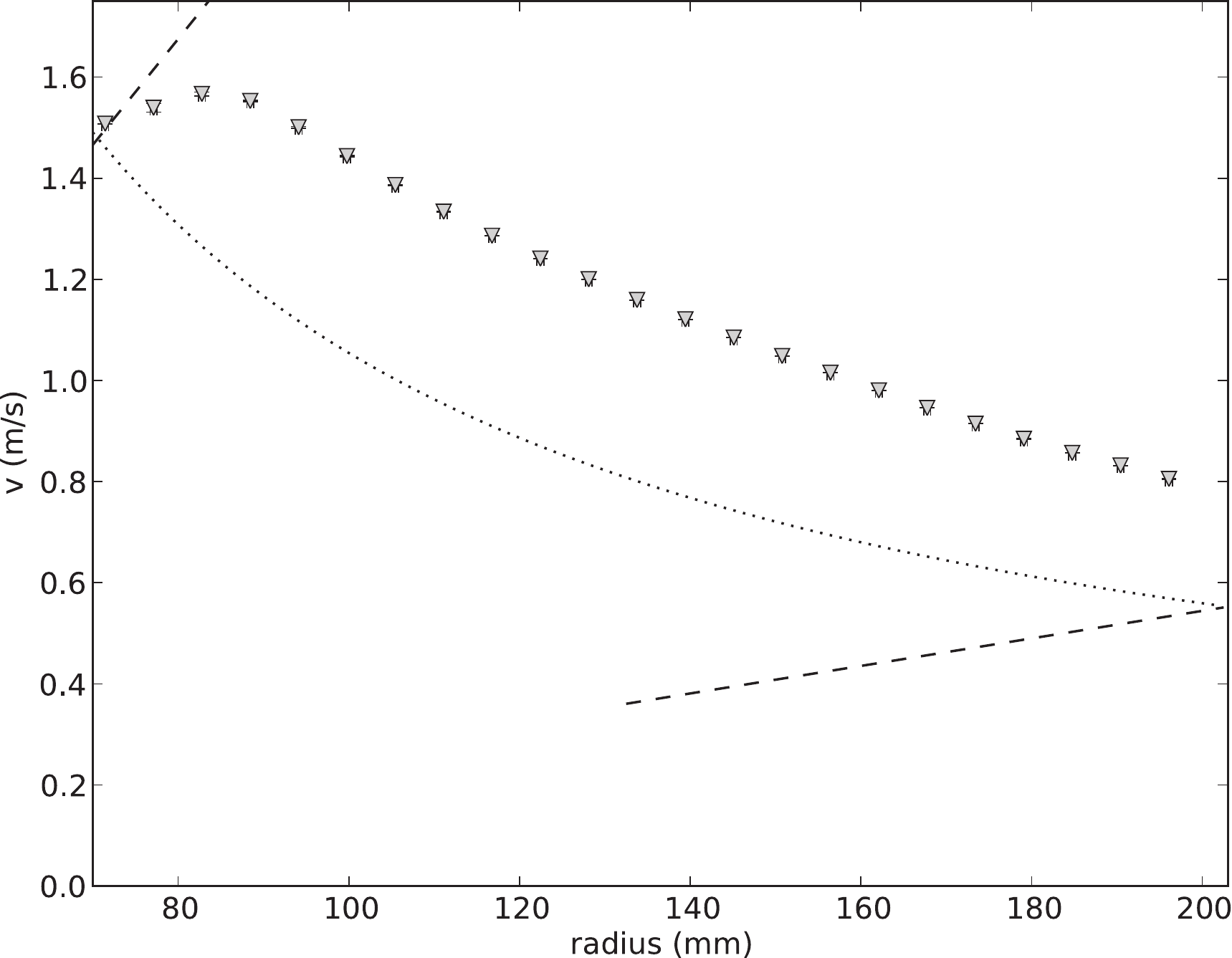} 
   \caption{Split end-cap operation at $Re=3\times10^5$. The dotted line is the ideal Couette profile. The dashed lines are the solid-body speeds of the end rings. Based on the cylinder speeds, the flow should be stable by the Rayleigh criterion. Our $beta$ measurement indicates that the bulk flow is in a turbulent state. Only Rayleigh-unstable flows with the outer cylinder at rest were observed to have greater levels of transport.}
   \label{fig:split_5}
\end{figure}

As a final note, we briefly remark on a new paper \citep{Paoletti:2011} that we were informed just before our planned submission of the present paper.
The authors presented a $\beta$ value of $(1.7 \pm 0.2) \times 10^{-5}$ inferred from torque measurements in quasi-Keplerian flows at similar Reynolds numbers to ours. Since this $\beta$ value is significantly above ours from direct measurements of Reynolds stress, it was claimed that a turbulent state is seen. We point out, however, that the large $\beta$ values measured there are subject to the interpretation based on the Ekman effects caused by
axial end caps corotating with the outer cylinder. In fact, their reported $\beta$ value is only by a factor of 2 larger than that of our QK(Ekman) flow (see Figure \ref{fig:beta_plot}), which has a reduced Ekman circulation due to a disrupted boundary layer despite the small aspect ratio \citep{Schartman:2009}.  \cite{Richard:2001} reported significant Ekman effects in quasi-Keplerian flows with corotating end caps (and also in the Split configuration) as indicated
by large deviations of the measured velocity profiles from the ideal Couette solution (see his Figure 4.2). Velocity profiles are unreported in \cite{Paoletti:2011} 
but their aspect ratio of 11.47 is smaller than $\sim$ 25 in \cite{Richard:2001}, and thus should be subject to stronger Ekman circulations. This latter point is consistent with recent results from numerical simulations \citep{Avila:2012} 
which were performed at $Re$ significantly below experimental values. The reported numerical results also showed turbulence in the bulk flow in our experimental geometry at low $Re$. However, these results do not necessarily contradict our data reported in the present paper: the measured $\beta$ values (e.g. the point A in Figure~\ref{fig:beta_plot}) 
are significantly larger at lower $Re$ when using glycerol mix even with optimal axial boundary conditions. We interpret this to mean that the residual unsteady secondary circulation penetrates deeper into the bulk flow at lower $Re$, but the detailed $Re$ dependence of secondary circulation is subject to future studies.
In any case, a quiescent flow, like quasi-Keplerian flows, is more likely to be delicate (than {\it e.g.} turbulent flows), 
and thus prone to be perturbed by imperfect axial boundary conditions even at large aspect ratios.

\begin{acknowledgements}
We acknowledge the support from the U.S. Department of Energy's Office of Sciences - Fusion Energy Sciences Program through
the contract number DE-AC02-09CH11466, the support from the U.S. National Science Foundation under grant numbers AST-0607472 and PHY-0821899, and the support from the U.S. National Aeronautics and Space Administration (NASA) under grant numbers APRA08-0066 and ATP06-35. We also appreciate comments and suggestions by referees.
\end{acknowledgements}

\bibliographystyle{aa}
\bibliography{Thesis}

\newpage

\end{document}